\documentclass[epj]{svjour}
\usepackage{latexsym}
\usepackage{url}
\usepackage{graphicx}
\usepackage{bm}   
\usepackage{amssymb}
\usepackage{amsmath}
\usepackage{amsfonts}
\usepackage{mathptmx}
\usepackage{dcolumn}                 
\usepackage{color}

\newcommand{\al}{\alpha}
\newcommand{\be}{\beta}

\newcommand{\de}{\delta}
\newcommand{\De}{\Delta}
\newcommand{\ep}{\varepsilon}

\newcommand{\la}{\lambda}

\renewcommand{\th}{\theta}   

\newcommand{\beq}{\begin{equation}}
\newcommand{\eeq}{\end{equation}}
\newcommand{\ba}{\begin{array}}
\newcommand{\ea}{\end{array}}
\newcommand{\bea}{\begin{eqnarray}}
\newcommand{\eea}{\end{eqnarray}}
\newcommand{\bi}{\begin{itemize}}  
\newcommand{\ei}{\end{itemize}}
\newcommand{\ben}{\begin{enumerate}} 
\newcommand{\een}{\end{enumerate}}
\newcommand{\bc}{\begin{center}}
\newcommand{\ec}{\end{center}}

\newcommand{\dsp}{\displaystyle}
\newcommand\eqn[1]{(\ref{#1})}      

\newcommand{\ee}[1]{\times 10^{#1}}

\newcommand{\km}{{\rm km}}

\newcommand{\ptrans}{p_{\rm trans}}
\newcommand{\ntrans}{n_{\rm trans}}
\newcommand{\etrans}{\varepsilon_{\rm trans}}

\newcommand{\Msolar}{{\rm M}_{\odot}}
\newcommand{\pcent}{p_{\rm cent}}

\newcommand{\cQM}{{c^{\phantom{1}}_{\rm QM}}}

\newcommand{\cQMsq}{c^2_{\rm QM}}

\newcommand{\destab}{\De\ep_{\rm crit}}
\newcommand{\Mmax}{M_{\rm max}}
\newcommand{\Rtyp}{R_{1.4}}

\def\bea{\begin{eqnarray}}
\def\eea{\end{eqnarray}}
\def\be{\begin{equation}}
\def\ee{\end{equation}}

\def\beq{\begin{equation}}
\def\eeq{\end{equation}}
\def\bar{\begin{array}[b]}
\def\barc{\begin{array}}
\def\bart{\begin{array}[t]}
\def\ear{\end{array}}

\usepackage[normalem]{ulem}  

\begin{document}
\title{Characteristics of hybrid
compact stars with a sharp hadron-quark interface}
\author{Mark G. Alford \and Sophia Han 
\thanks{\emph{e-mail:} jhan@physics.wustl.edu }%
}                     
%
%
\institute{Physics Department, Washington University, St. Louis, Missouri 63130, USA}
\date{Received: 3 August 2015 / Revised: 20 January 2016; Published online: 22 March 2016}
%
\abstract{
We describe two aspects of the physics of hybrid stars that
have a sharp interface
between a core of quark matter and a mantle of nuclear matter. Firstly, we analyze the mass-radius relation.
We describe a generic ``Constant-Speed-of-Sound'' (CSS) parameterization of the quark matter equation of state (EoS), in which the speed of sound is independent of density. In terms of the three parameters of the CSS EoS we obtain the phase diagram of possible forms of the hybrid star mass-radius relation, and we
show how observational constraints on the maximum mass and typical radius
of neutron stars can be expressed as constraints on the CSS parameters. Secondly, we propose a mechanism for the damping of density oscillations, 
including r-modes, in hybrid stars with a sharp interface.
The dissipation arises from the periodic conversion
between quark matter and nuclear matter induced by the
pressure oscillations in the star. We find the damping grows
nonlinearly with the amplitude of the oscillation and is powerful enough to saturate an r-mode at very low saturation amplitude, of order $10^{-10}$, which
is compatible with currently-available
observations of neutron star spin frequencies and temperatures.
} 
\maketitle
\section{Introduction}
\label{sec:intro}

One of the major goals of nuclear physics is to establish the nature and
properties of matter at nuclear density and beyond.
Experimental constraints on the properties of ultra-high-density matter
come from neutron stars, where gravity compresses matter to the
highest densities found in nature. One hypothesis is that there
is a phase transition from nuclear matter to quark matter, and that
some stars, called ``hybrid stars'', contain cores of quark matter.

We will assume that in hybrid stars 
the core of quark matter and the mantle of nuclear matter 
are separated by a sharp interface (Maxwell
construction) as opposed to a mixed phase (Gibbs construction).
This assumption is valid if the nuclear-quark phase transition is
first order, and the surface tension of the interface is high enough.
This is a possible scenario, given the uncertainties in the value
of the surface tension  \cite{Alford:2001zr,Palhares:2010be,Pinto:2012aq}.
For analysis of generic equations of state that 
continuously interpolate between the phases to model mixing or percolation,
see Refs.~\cite{Macher:2004vw,Masuda:2012ed}.

In this paper we discuss two aspects of the physics of such  hybrid stars,
each of which is relevant to a different set of observable properties of the 
star.

Firstly, in Secs.~\ref{sec:CSS-intro} to \ref{sec:CSS-constraints},
we analyze the mass-radius relation of hybrid stars, using a 
a generic parameterization of the quark matter equation of state (EoS).
We obtain the phase diagram of possible forms of the hybrid star 
mass-radius relation, and we
show how observational constraints on the maximum mass and typical radius
of hybrid stars can be expressed as constraints on the parameters of the 
quark matter EoS.

Secondly, in Sec.~\ref{sec:phase-conversion},
we propose a mechanism for the damping of density oscillations
in hybrid stars with a sharp interface. Such damping is important for
the occurrence and saturation of r-modes 
\cite{Andersson:1997xt,Andersson:2000mf}, which if undamped can
spin down a star in days via emission of gravitational radiation.
The relevant observables in this case are the spin frequency,
spindown rate, and temperature of the star.
Our proposed damping mechanism
arises from the periodic conversion
between quark matter and nuclear matter induced by the
pressure oscillations in the star. We find the damping grows
nonlinearly with the amplitude of the oscillation and is powerful enough to saturate an r-mode at very low saturation amplitude, of order $10^{-10}$, which
is compatible with currently-available
observations of neutron star spin frequencies and temperatures.

\section{The CSS parameterization}
\label{sec:CSS-intro}

\begin{figure}[htb]
\parbox{\hsize}{
\includegraphics[width=\hsize]{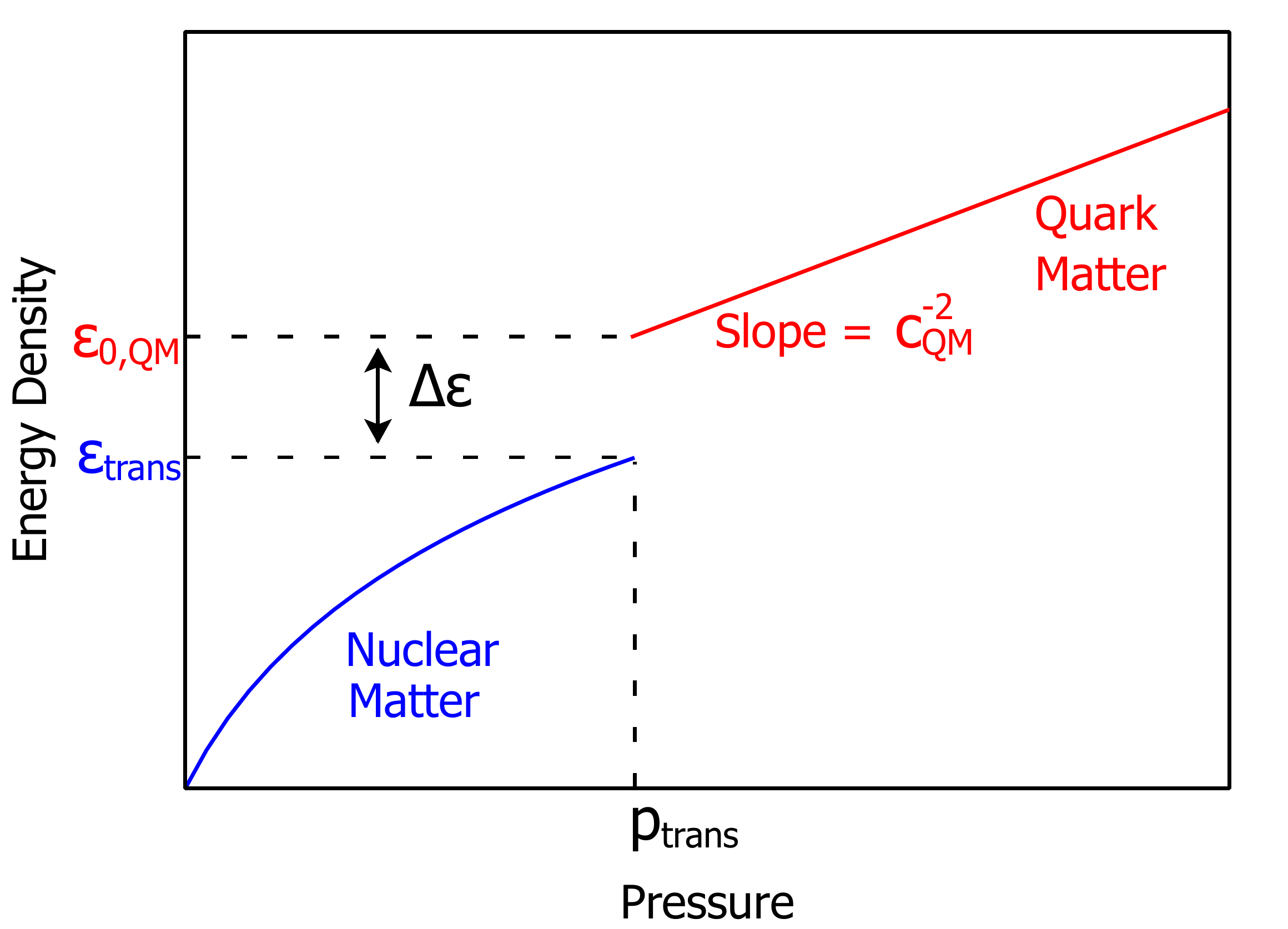}
}
\caption{Equation of state $\ep(p)$
for dense matter. The quark matter EoS is specified by 
the transition pressure  $\ptrans$, the
energy density discontinuity
$\De\ep$, and the speed of sound in quark matter $\protect\cQM$
(assumed density-independent). 
}\label{fig:EoS-schematic}
\end{figure}

The equation of state (EoS) of quark matter at the densities and temperatures
relevant to neutron stars has not yet been calculated from first
principles. There are many models available, each based on some simplification
of the real physics, and each with their own parameters.  It is therefore
useful to have a general parameterization of the EoS which can be used as a
generic language for relating different models to each other and for
expressing experimental constraints in model-independent terms.
We will use the ``Constant-Speed-of-Sound'' (CSS) parameterization 
\cite{Alford:2013aca,Zdunik:2012dj,Chamel:2012ea}, which
can be viewed as the lowest-order terms of a Taylor expansion of the
quark matter EoS about the transition pressure $\ptrans$.
As illustrated in Fig.~\ref{fig:EoS-schematic}, one can then write
the high-density EoS in terms of three
parameters: the pressure $\ptrans$ of the transition, 
the discontinuity in energy density $\De\ep$ at the transition,
and  the speed of sound $\cQM$ in the high-density phase.
For a given nuclear matter EoS $\ep_{\rm NM}(p)$, the full CSS EoS is then
\beq
\ep(p) = \left\{\!
\begin{array}{ll}
\ep_{\rm NM}(p) & p<\ptrans \\
\ep_{\rm NM}(\ptrans)+\De\ep+c_{\rm QM}^{-2} (p-\ptrans) & p>\ptrans
\end{array}
\right.\ ,
\label{eqn:EoSqm1}
\eeq
where $\ep_{\rm NM}(p)$ is the nuclear matter equation of state.

The assumption of a density-independent speed of sound is 
accurate for a
large class of models of quark matter, including many Nambu--Jona-Lasinio models \cite{Zdunik:2012dj,Agrawal:2010er,Bonanno:2011ch,Lastowiecki:2011hh},
the perturbative quark
matter EoS \cite{Kurkela:2010yk},
and the quartic polynomial parameterization \cite{Alford:2004pf}.
For details, see Ref.~\cite{Alford:2015dpa}.
It is noticeable that models based on relativistic quarks
tend to have $\cQMsq\approx$ close to 1/3, which is the value for
systems with conformal symmetry, and it has been conjectured that 
there is a fundamental bound $\cQMsq<1/3$ \cite{Bedaque:2014sqa}, although some models violate that bound, e.g.~\cite{Kojo:2014rca,Benic:2014iaa} or \cite{Lastowiecki:2011hh} (parameterized in \cite{Zdunik:2012dj}).

\begin{figure*}[htb]
\parbox{0.25\hsize}{
\centerline{\large ``Absent''}
\includegraphics[width=\hsize]{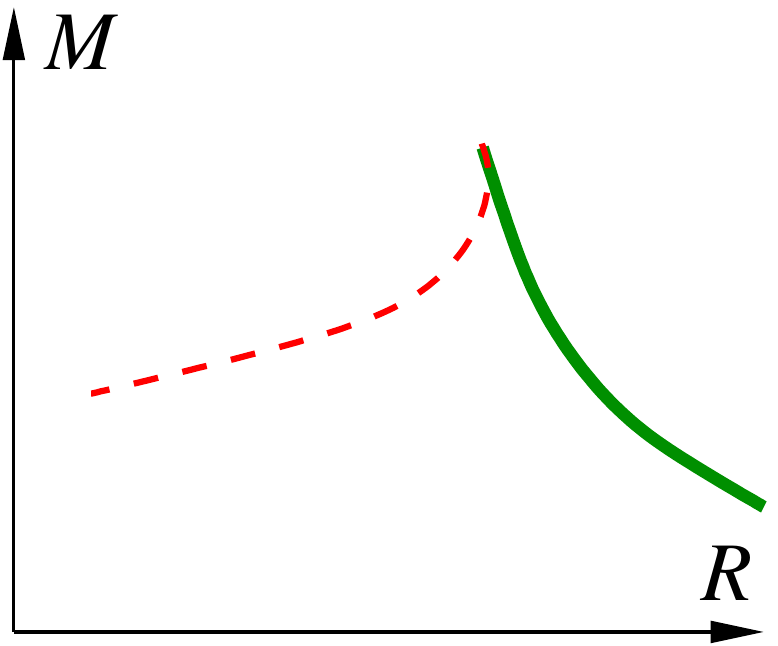}
\bc (a) \ec
}\parbox{0.25\hsize}{
\centerline{\large ``Both''}
\includegraphics[width=\hsize]{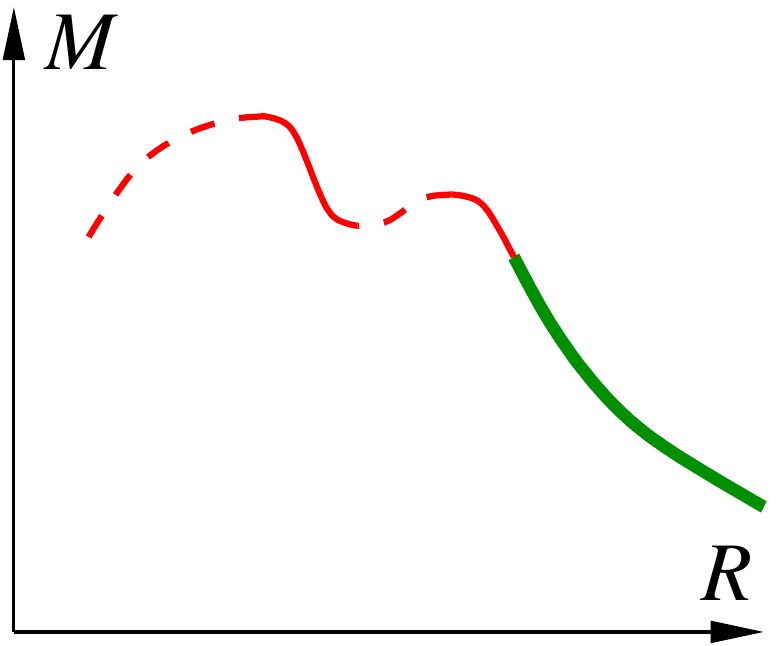}
\bc (b) \ec
}\parbox{0.25\hsize}{
\centerline{\large ``Connected''}
\includegraphics[width=\hsize]{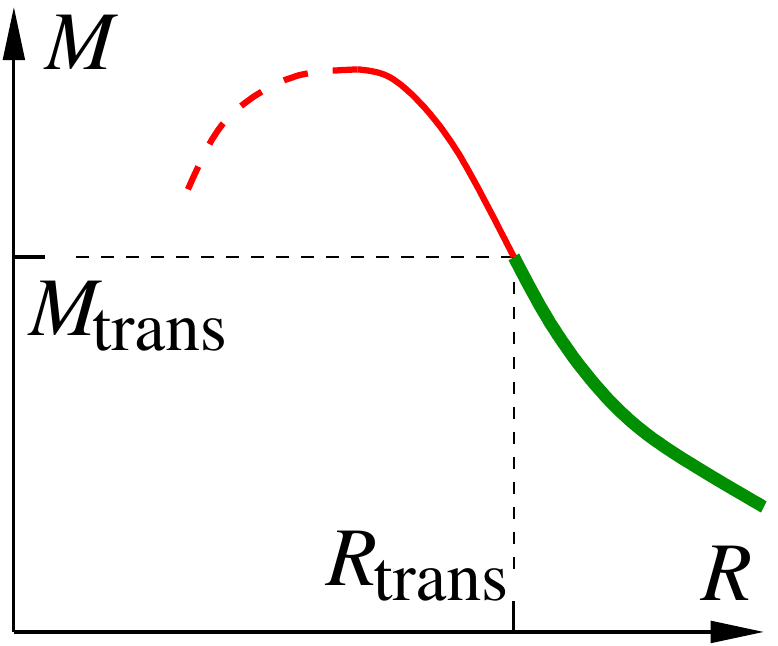}
\bc (c) \ec
}\parbox{0.25\hsize}{
\centerline{\large ``Disconnected''}
\includegraphics[width=\hsize]{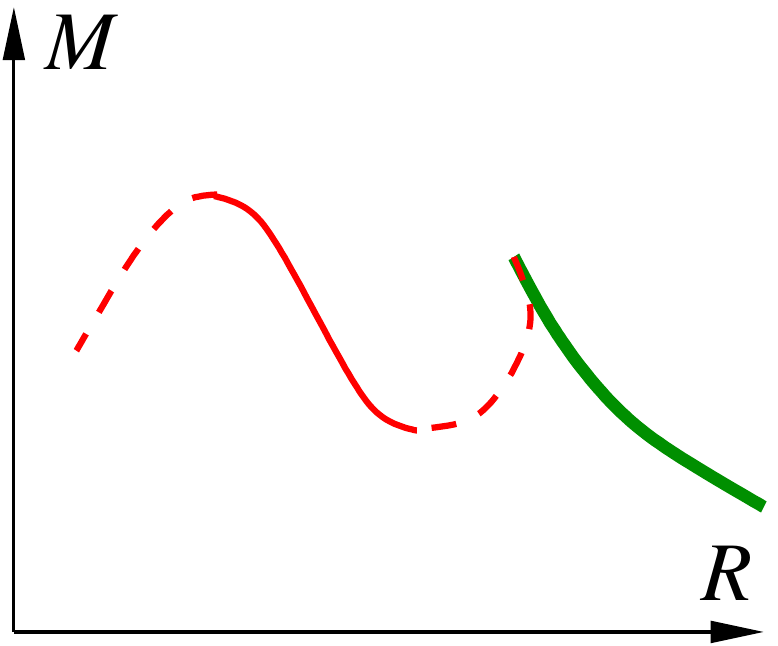}
\bc (d) \ec
}
\caption{(Color online) Schematic form of possible mass-radius relations for hybrid stars.
The thick (green) line is the hadronic branch. Thin solid (red) lines
are stable hybrid stars; thin dashed (red) lines are unstable hybrid stars.
In (a) the hybrid branch is absent. In (c) there is a connected branch.
In (d) there is a disconnected branch. In (b) there are both types of branch.
In realistic neutron star $M(R)$ curves
the cusp that occurs in cases (a) and (d) is much smaller and
harder to see \cite{Haensel:1983,Lindblom:1998dp}.
}
\label{fig:MR-De}
\end{figure*}

\begin{table}[htb]
\begin{center}
\begin{tabular}{c c@{\quad} c}
\hline
property & BHF, \, Av${}_{18}$  & DBHF, \\
         & + UVIX~TBF           & Bonn~A \\
\hline
saturation baryon density $n_0 \rm (fm^{-3})$   & 0.16 & 0.18  \\
binding energy/baryon $E/A$ (MeV)     & -15.98  &  -16.15  \\
compressibility $K_0$ (MeV)           & 212.4 &  230  \\
symmetry energy $S_0 $ (MeV)          & 31.9  &  34.4  \\
$L= 3n_0(\partial S_0/\partial n)|_{n_0}$ (MeV)    & 52.9  &  69.4 \\
maximum mass of star ($\Msolar$)          & 2.03 & 2.31 \\
radius of the heaviest star (km)  & 9.92 &11.26 \\
radius of $M=1.4\,\Msolar$ star (km)  & 11.77 &13.41 \\
\hline
\end{tabular}
\end{center}
\caption{Calculated properties of symmetric nuclear matter for the BHF and DBHF
nuclear equations of state used here. BHF is softer, DBHF is stiffer.}
\label{tab:EoS}
\end{table}  

At pressures below $\ptrans$ we use
two examples of nuclear EoS: Dirac-Brueckner-Hartree-Fock (DBHF) \cite{GrossBoelting:1998jg} and Brueck-
ner-Hartree-Fock (BHF) \cite{Taranto:2013gya}. Some of the properties of BHF and DBHF are summarized in
Table~\ref{tab:EoS}, where $L$ is related to the derivative of the
symmetry energy $S_0$ with respect to density at the nuclear saturation
density, $L= 3n_0(\partial S_0/\partial n)|_{n_0}$. BHF is a softer equation of state, with a lower value of $L$ and lower
pressure at a given energy density.
DBHF is a stiffer EoS, with higher pressure at a given energy density. 
It yields neutron stars that are
larger, and can reach a higher maximum mass. With CSS parametrization for quark matter and realistic nuclear matter EoSs, we study in detail what possible forms of hybrid star mass-radius relations are and how they rely on the high-density phase parameter values.

In Sec.~\ref{sec:CSS-phase-diagram} we 
show how the CSS parameterization is
constrained by observables such as the
maximum mass $\Mmax$, the radius of a maximum-mass star, and the radius $\Rtyp$
of a star of mass $1.4\,\Msolar$, based on BHF and DBHF nuclear matter EoSs.

\section{Generic conditions for stable hybrid stars}
\label{sec:CSS-phase-diagram}

\begin{figure*}[htb]
\parbox{0.5\hsize}{
\includegraphics[width=\hsize]{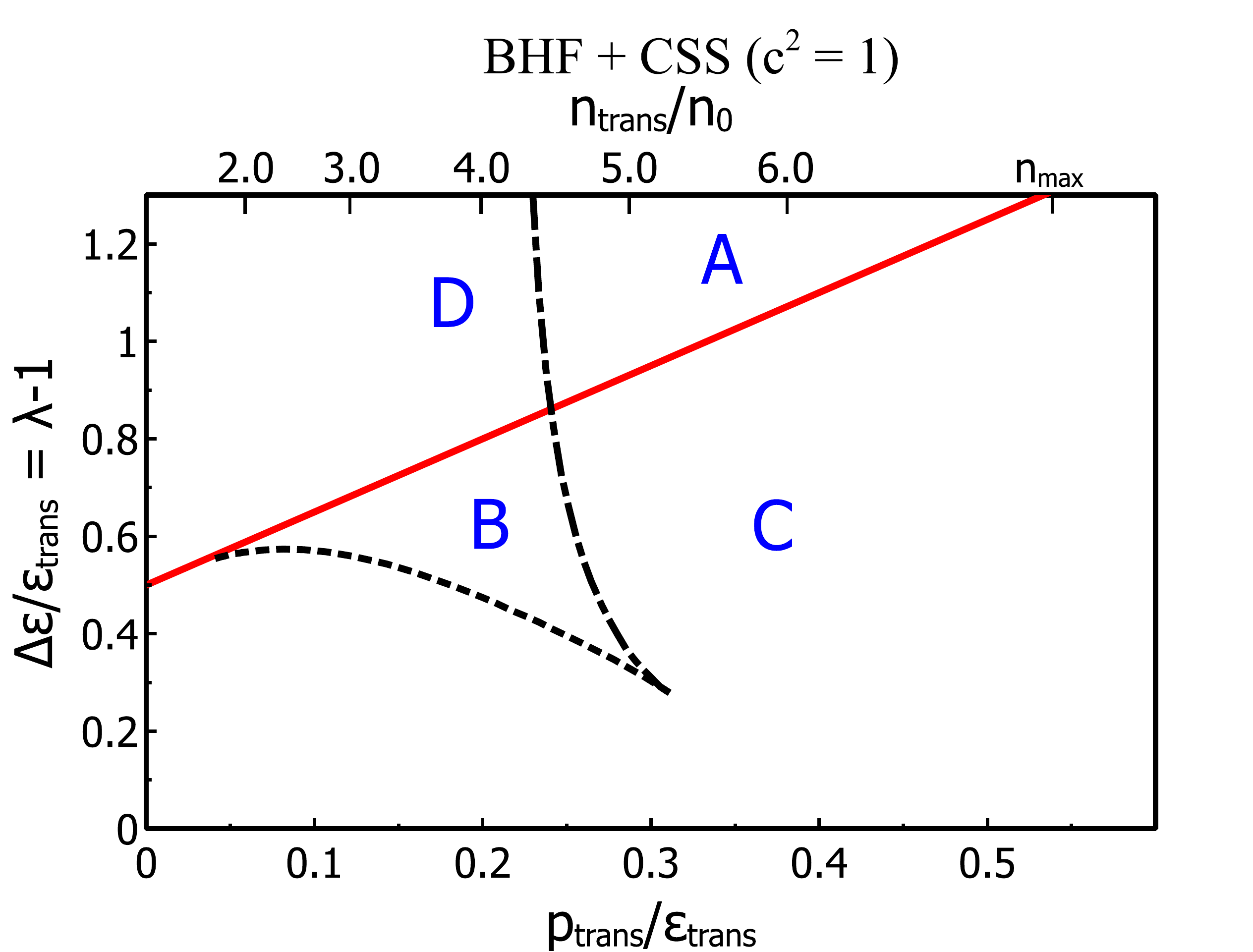}\\[2ex]
}\parbox{0.5\hsize}{
\smallskip
\includegraphics[width=0.9\hsize]{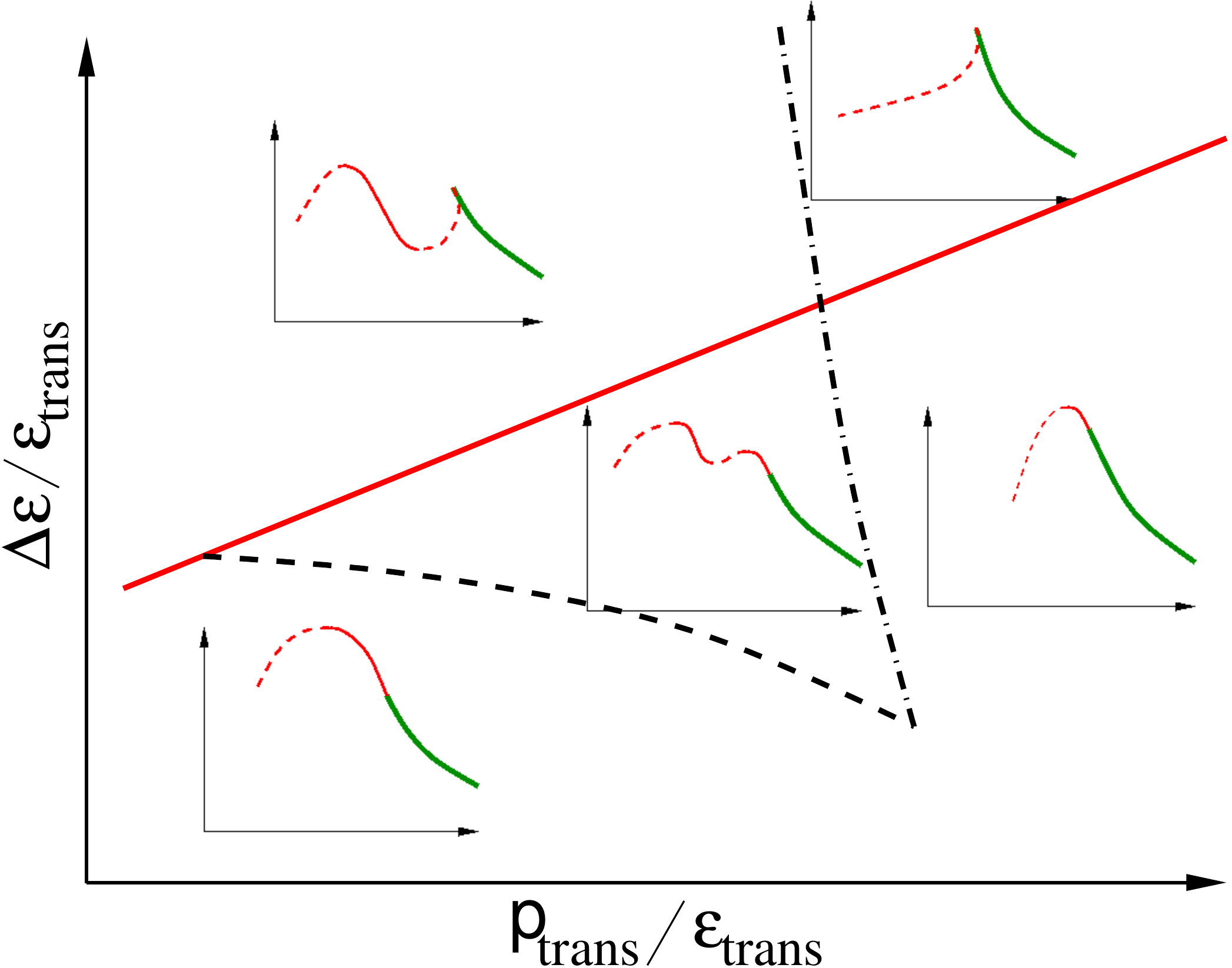}\\[2ex]
}
\caption{(Color online) Phase diagram for hybrid star branches in the
mass-radius relation of compact stars.
}
\label{fig:phase-diag-HLPS}
\end{figure*}

\begin{figure*}
\parbox{0.5\hsize}{
\vspace{-2ex}
\includegraphics[width=\hsize]{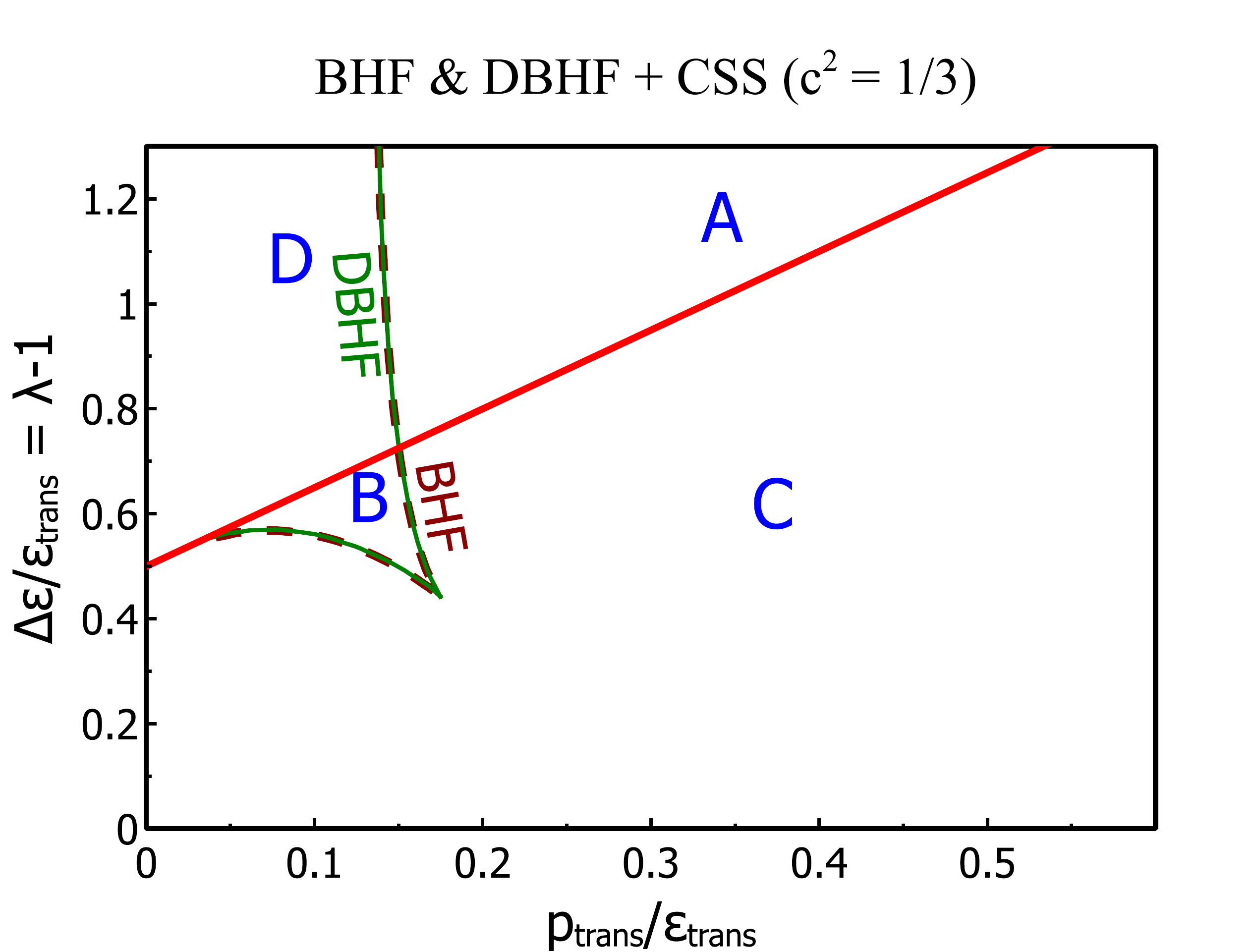}\\[2ex]
}\parbox{0.5\hsize}{
\vspace{-2ex}
\includegraphics[width=\hsize]{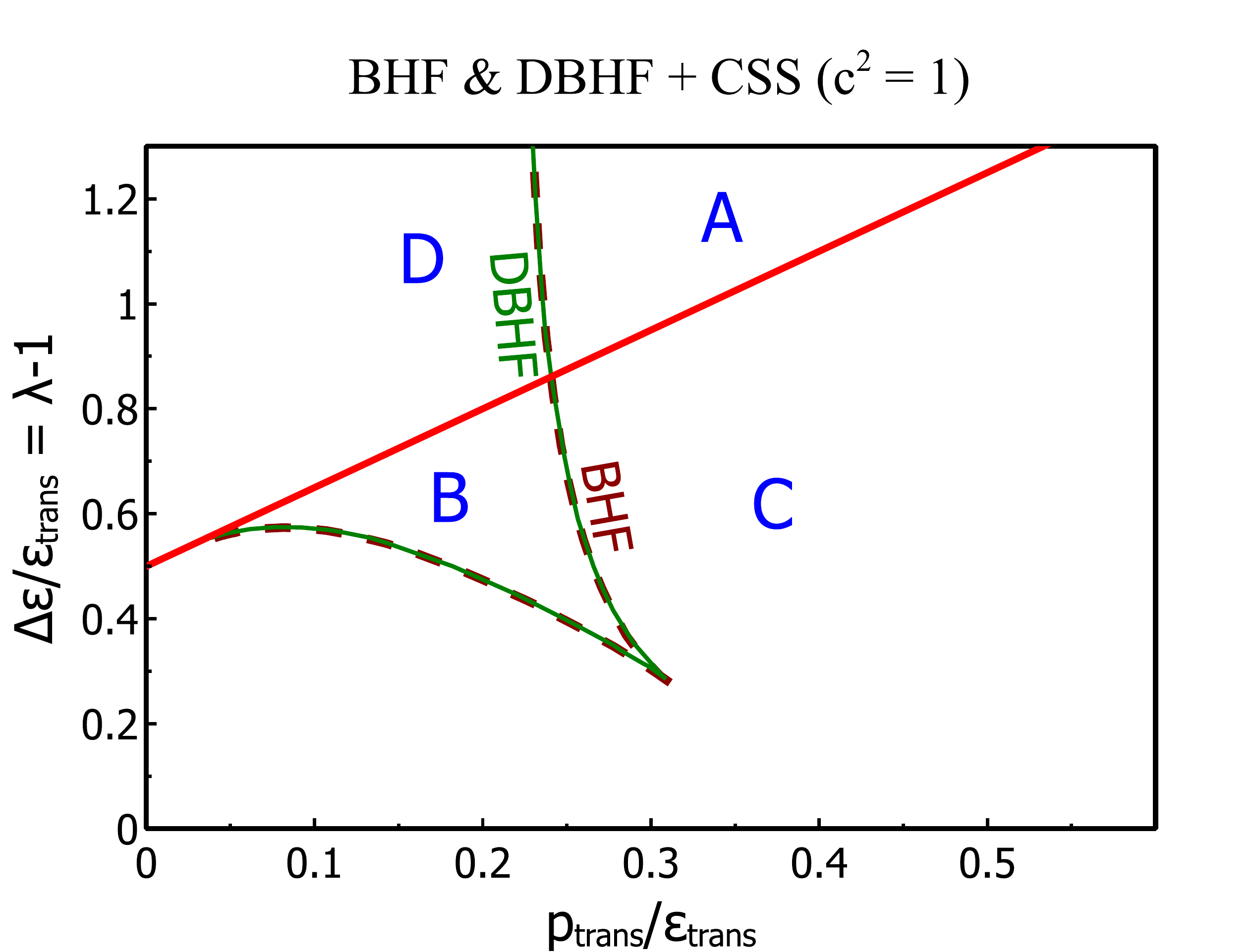}\\[2ex]
}
\caption{(Color online) Phase diagram like  Fig.~\ref{fig:phase-diag-HLPS}, showing that
the phase boundaries are very insensitive to changes in the nuclear 
matter EoS, but they are affected by varying the quark matter speed of sound.
Note that, for a given pressure, BHF and DBHF have very
different baryon densities.
}
\label{fig:phase-diag-both}
\end{figure*} 

A compact star will be stable as
long as the mass $M$ of the star is an increasing function of the
central pressure $\pcent$ \cite{Bardeen:1966}. By performing an expansion in powers of the size of the quark
matter core, it 
has been shown \cite{Haensel:1983,Lindblom:1998dp,Seidov:1971} that the appearance of a quark matter core when $\pcent$ rises above $\ptrans$ will destabilize the star only if the energy density discontinuity $\De\ep$
is too large, specifically if $\De\ep >\destab$ where
\beq
\frac{\destab}{\etrans} = \frac{1}{2} + \frac{3}{2}\frac{\ptrans}{\etrans} \ .
\label{eqn:stability}
\eeq
(This is $\la_{\rm crit}-1$ in the notation of Ref.~\cite{Haensel:1983}.)

In Fig.~\ref{fig:MR-De},  panels (b) and (c) show possible forms
of $M(R)$ for $\De\ep<\destab$, where a stable connected hybrid
branch continues on from the hadronic branch of the $M(R)$ relation.
Panels (a) and (d) show possible forms
for $\De\ep>\destab$, where there is no stable 
connected branch, although in (d) there is a disconnected one. In Figs.~\ref{fig:MR-De}~(b) and (d), 
we illustrate how, as noted in Ref.~\cite{Haensel:1983}, a second,
disconnected, branch of stable hybrid stars can arise.
The disconnected branch is a ``third family''
\cite{Glendenning:2000,Schertler:2000} of compact stars besides neutron stars
and white dwarfs.
In $M(R)$ curves for realistic neutron star equations
of state, the cusp that occurs where an unstable hybrid branch meets the
hadronic branch is exaggerated in Fig.~\ref{fig:MR-De}. In 
$M(R)$ plots for realistic EoSs the cusp is invisibly small,
covering a range in $M$ of less than one part in a thousand.

\subsection{Phase diagram at fixed $\protect\cQM$}

In Fig.~\ref{fig:phase-diag-HLPS} we plot a ``phase diagram'' for hybrid stars
showing how the form of the $M(R)$ relation depends on
the CSS parameters $\ptrans/\etrans$ and $\De\ep/\etrans$
at fixed $\cQMsq$ (see Ref.~\cite{Alford:2013aca,Alford:2015dpa}).
The left panel of Fig.~\ref{fig:phase-diag-HLPS}
is the result of calculations for the BHF nuclear matter EoS and quark matter with $\cQMsq=1$.
The right panel is a schematic showing the form of the mass-radius
relation in each region of the diagram. The regions correspond to
different geometries of the hybrid branch displayed
in Fig.~\ref{fig:MR-De}:
A=``Absent'', C=``Connected'', D=``Disconnected'', B=``Both''
(connected and disconnected).

The solid straight (red) line is $\destab$ from Eq.~\eqn{eqn:stability}.
Below the line in regions B and C, there is a hybrid star
branch connected to the nuclear star branch. Above the line in regions A and D, there is no connected
hybrid star branch. In regions B and D there is
a disconnected hybrid star branch.

The roughly vertical dash-dotted curve in  Fig.~\ref{fig:phase-diag-HLPS}
marks a transition where the disconnected branch of hybrid
stars appears/disappears. When one crosses this line from the right,
going from region A to D by decreasing the nuclear/quark
transition density, a stationary point of inflection appears in $M(\pcent)$
at $\pcent>\ptrans$.
If one crosses from C to B then this point of inflection is at higher
central pressure than the existing maximum in $M(\pcent)$. This 
produces a stationary point of inflection 
in the $M(R)$ relation to the left of the existing maximum (if any).
After crossing the dash-dotted line the point of inflection
becomes a new maximum-minimum pair (the maximum being further from the
transition point), producing a disconnected branch of stable hybrid stars
in regions B and D.
Crossing the other way, by increasing the transition pressure,
the maximum and minimum that delimit the
disconnected branch merge and the branch disappears.

The roughly horizontal dashed curve
in Fig.~\ref{fig:phase-diag-HLPS} which separates region B and C 
marks a transition between mass-radius relations with one connected
hybrid star branch, and those with two hybrid star branches, one connected
and one disconnected. 
Crossing this line from below, by increasing
the energy density discontinuity, 
a stationary point of inflection in $M(\pcent)$
(or equivalently in $M(R)$) appears in the existing
connected hybrid branch.
Crossing in to region B, this point of inflection
becomes a new maximum-minimum pair, so the connected hybrid branch
is broken in to a smaller connected branch and a new disconnected branch.
The maximum of the old connected branch smoothly becomes the maximum
of the new disconnected branch.
If one crossed the dashed line in the 
opposite direction, from B to C, the
maximum closest to the transition point would approach the minimum
and they would annihilate, leaving only the more distant maximum.

Where the horizontal and vertical curves meet, the two maxima and the minimum
that are present in region B all merge to form a single flat maximum
where the first three derivatives of $M(R)$ are all zero.

\subsection{Varying $\protect\cQM$ and the nuclear EoS}

In Fig.~\ref{fig:phase-diag-both} we show the effects of varying the speed of sound and the
nuclear matter EoS.
Fig.~\ref{fig:phase-diag-both}\,(a) is the phase diagram with $\cQMsq=1/3$ (characteristic of very weakly interacting massless quarks),
and Fig.~\ref{fig:phase-diag-both}\,(b) is for $\cQMsq=1$ (the maximum value consistent with causality). In both panels
we show the phase diagram for BHF and DBHF nuclear matter EoSs. The 
straight line is independent of $\cQMsq$ and the detailed form
of the nuclear matter EoS.
The other phase boundaries, outlining the region where there is a
disconnected hybrid branch, are remarkably insensitive to the
details of the nuclear matter EoS but depend strongly on
the quark matter speed of sound.  For a given nuclear matter EoS
the hybrid branch
structure is determined by $\ptrans/\etrans$, $\De\ep/\etrans$, and
$\cQMsq$, so one could make a three-dimensional plot with $\cQMsq$ as
the third axis, but this figure adequately illustrates the dependence 
on $\cQMsq$. We will now discuss the physics behind the shape of the
phase boundaries.

\subsection{Physical understanding of the phase diagram}

The main feature of the phase diagram is that
a disconnected branch is present when
the transition density is sufficiently low, and the energy density discontinuity
is sufficiently high.
It occurs more readily if the speed of sound in quark matter is high.

When a very small quark matter core is present, its greater density creates
an additional gravitational pull on the nuclear mantle. If the pressure of the
core can counteract the extra pull, the star is stable
and there is a connected hybrid branch.
If the energy density jump is too great, the extra gravitational pull is too
strong, and the star becomes unstable when quark matter first appears. However, if
the energy density of the core rises slowly enough with increasing pressure
(i.e. if  $c^2=dp/d\ep$ is large enough), a larger core with
a higher central pressure may be able to sustain the weight of the nuclear
mantle above it. Region B, with connected and disconnected branches,
is more complicated and we do not have an intuitive explanation for it.

We can now understand why the
vertical line marking the B/C and D/A boundaries moves to the right as $\cQMsq$
increases. Since $c^2=dp/d\ep$, if $\cQMsq$
is larger then the energy density of the core rises 
more slowly with increasing pressure, which minimizes the tendency for
a large core to destabilize the star via its gravitational attraction.
Finally, we can see why that line has a slight negative slope: larger
$\De\ep$ makes the quark core heavier, increasing its pull on the nuclear 
mantle, and making the hybrid star more unstable against collapse.

\section{Constraining a generic high-density equation of state}
\label{sec:CSS-constraints}

\subsection{Maximum mass of hybrid stars}

\begin{figure*}[htb]
\parbox{0.5\hsize}{
\includegraphics[width=\hsize]{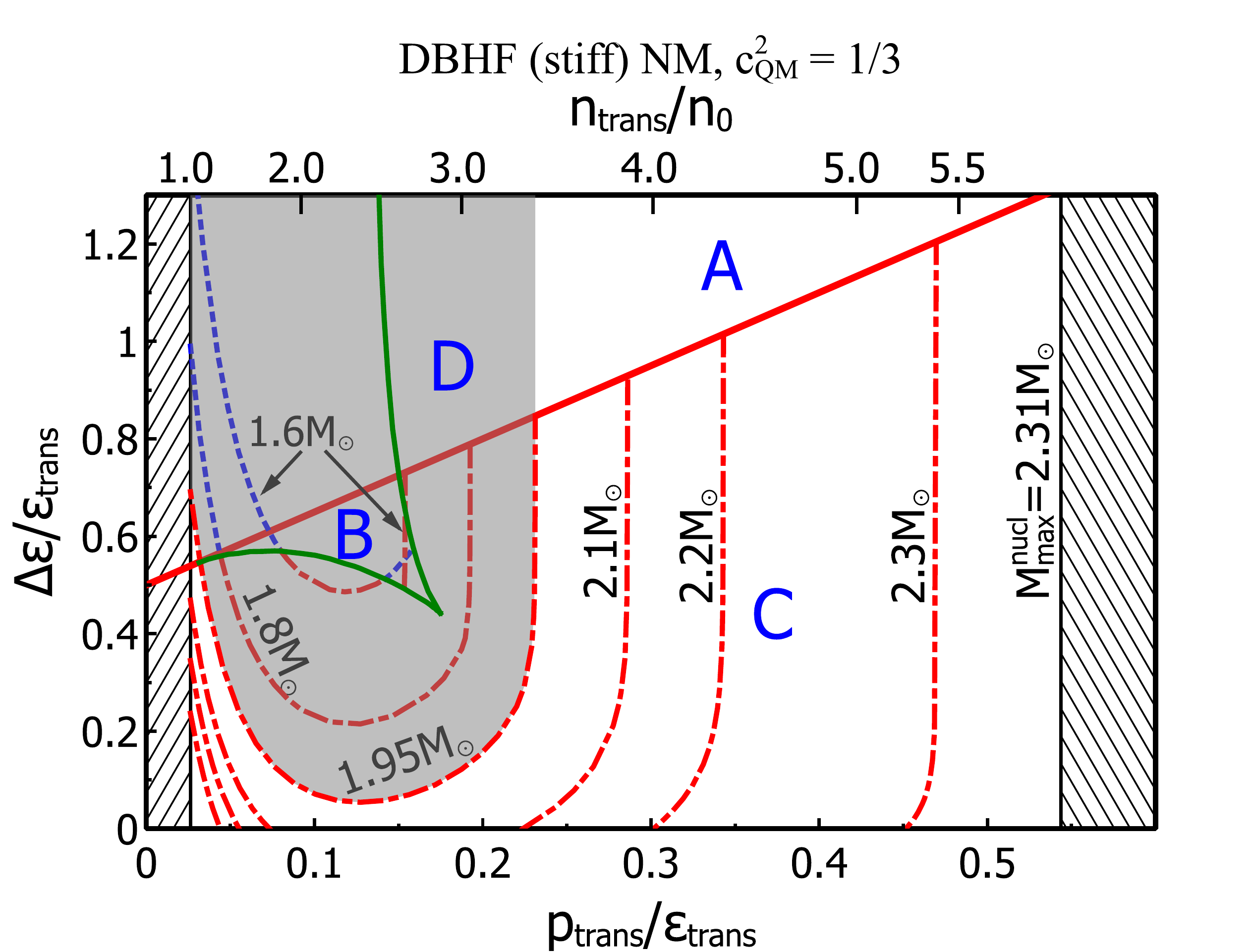}\\[2ex]
}\parbox{0.5\hsize}{
\includegraphics[width=\hsize]{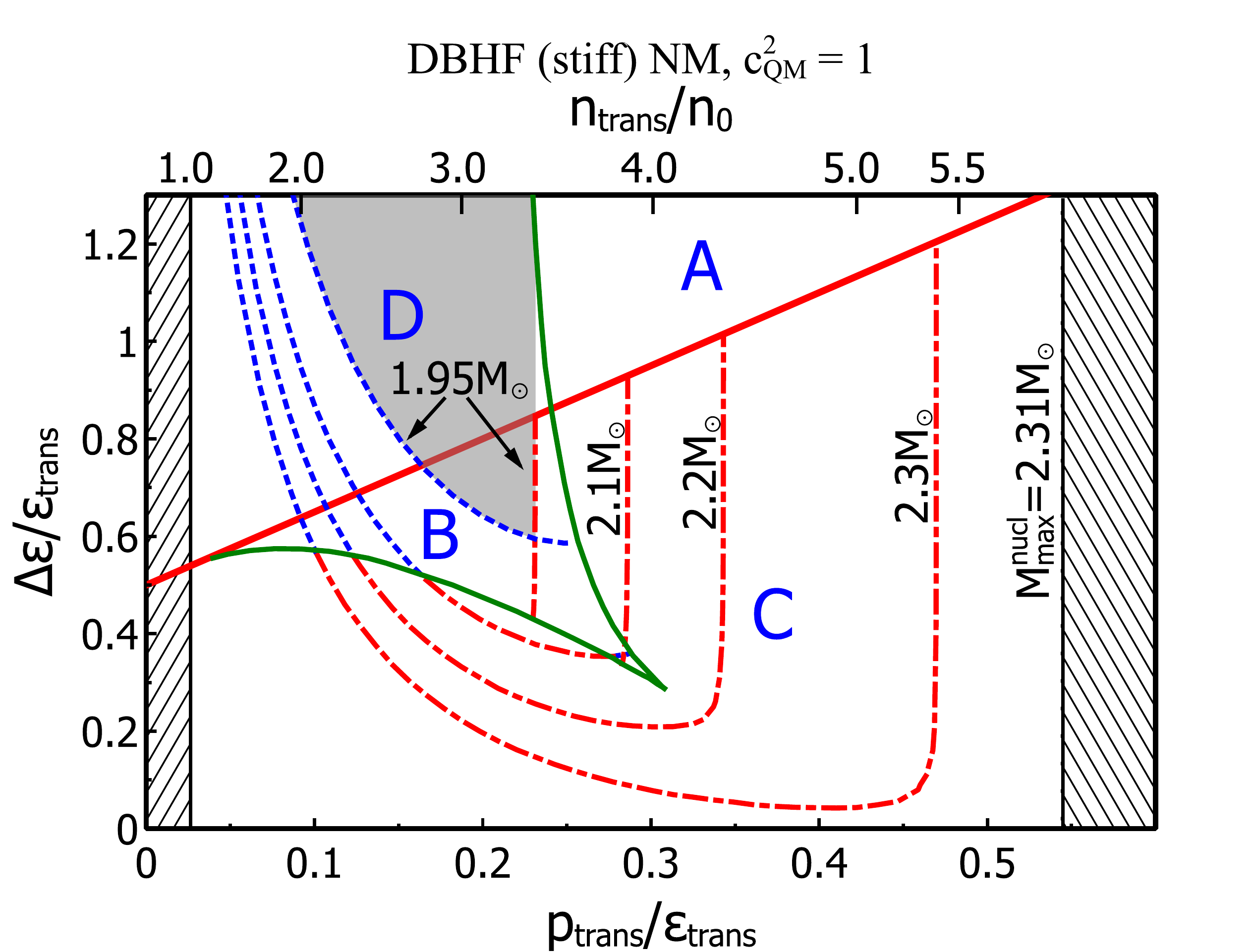}\\[2ex]
}\\[2ex]
\parbox{0.5\hsize}{
\includegraphics[width=\hsize]{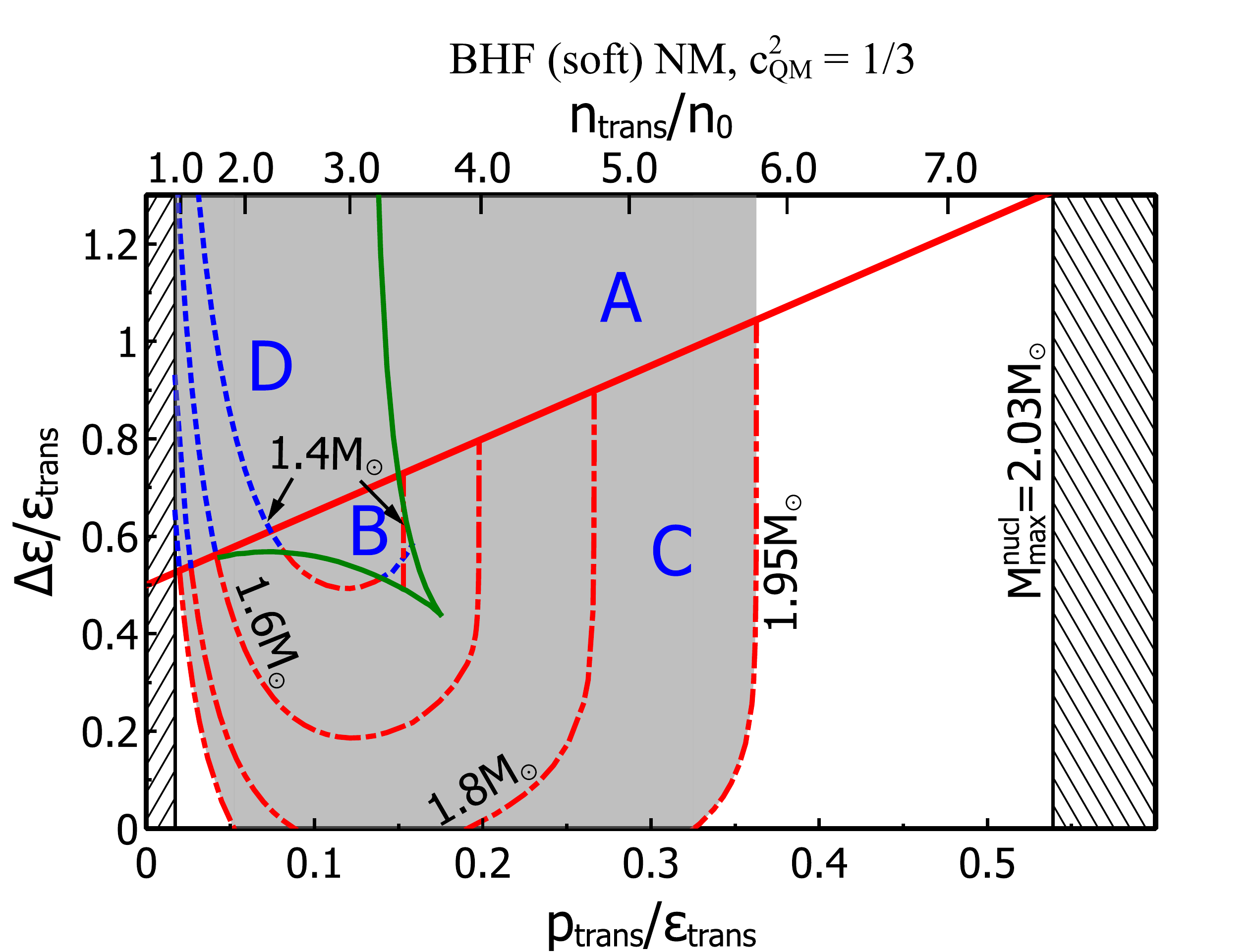}\\[2ex]
}\parbox{0.5\hsize}{
\includegraphics[width=\hsize]{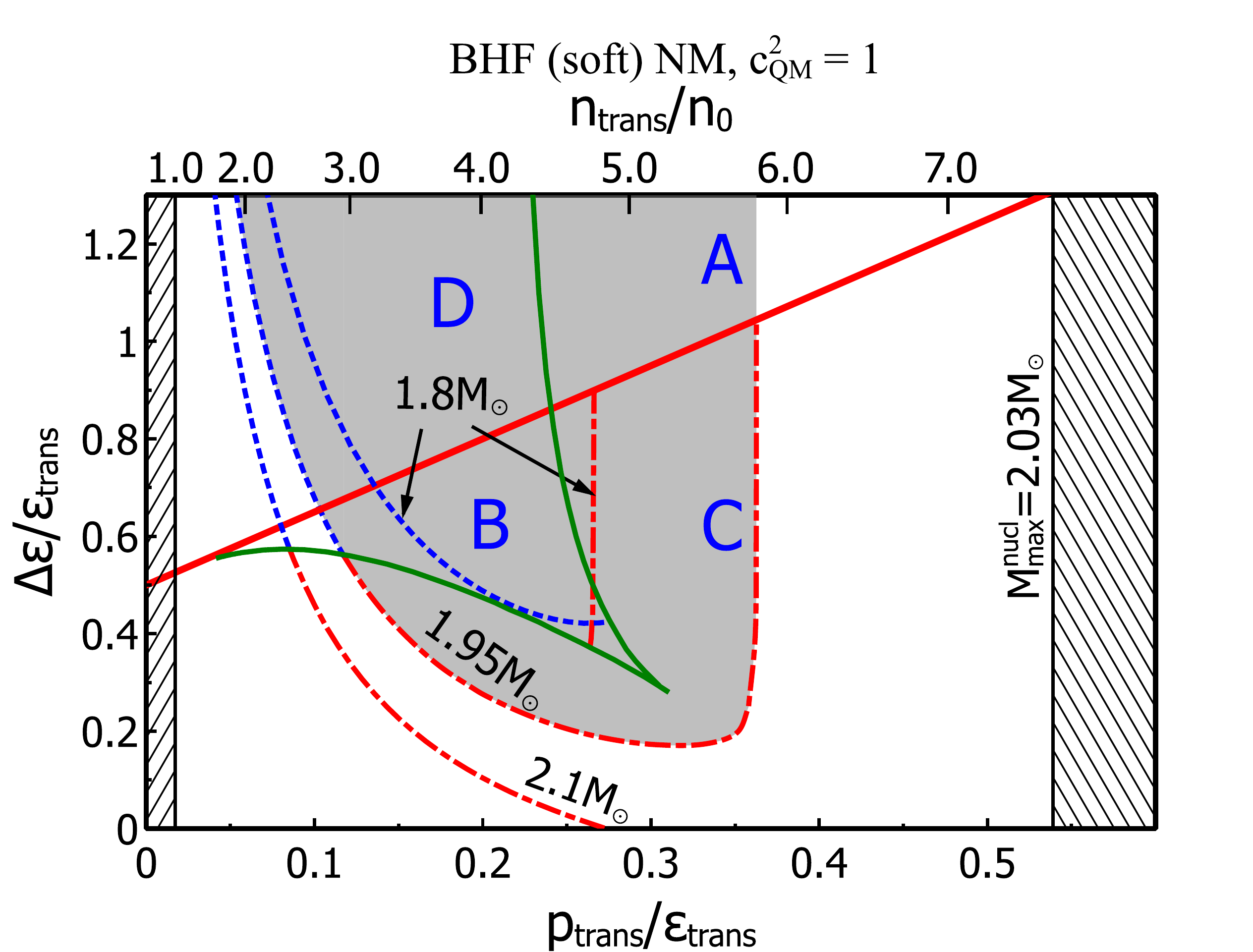}\\[2ex]
}
\caption{(Color online) Contour plots showing the maximum hybrid star mass as a function of the
CSS parameters of the high-density EoS. Each panel shows the dependence
on the CSS parameters $\ptrans/\etrans$ and $\De\ep/\etrans$. The left
plots are for  $\cQMsq=1/3$, and the right plots are for $\cQMsq=1$.
The top row is for a DHBF (stiff) nuclear matter EoS, and the bottom row
is for a BHF (soft) nuclear matter EoS.
The grey shaded region, where $\Mmax < 1.95\,\Msolar$,
is excluded by the observation of stars of mass $M\approx 2\,\Msolar$.
The hatched band at low density (where $\ntrans<n_0$) is
excluded because bulk nuclear matter would be metastable. The
hatched band at high density is excluded because the transition pressure
is above the central pressure of the heaviest stable hadronic star.
}
\label{fig:CSS-max-mass}
\end{figure*}

\begin{figure*}[htb]
\parbox{0.5\hsize}{
\includegraphics[width=\hsize]{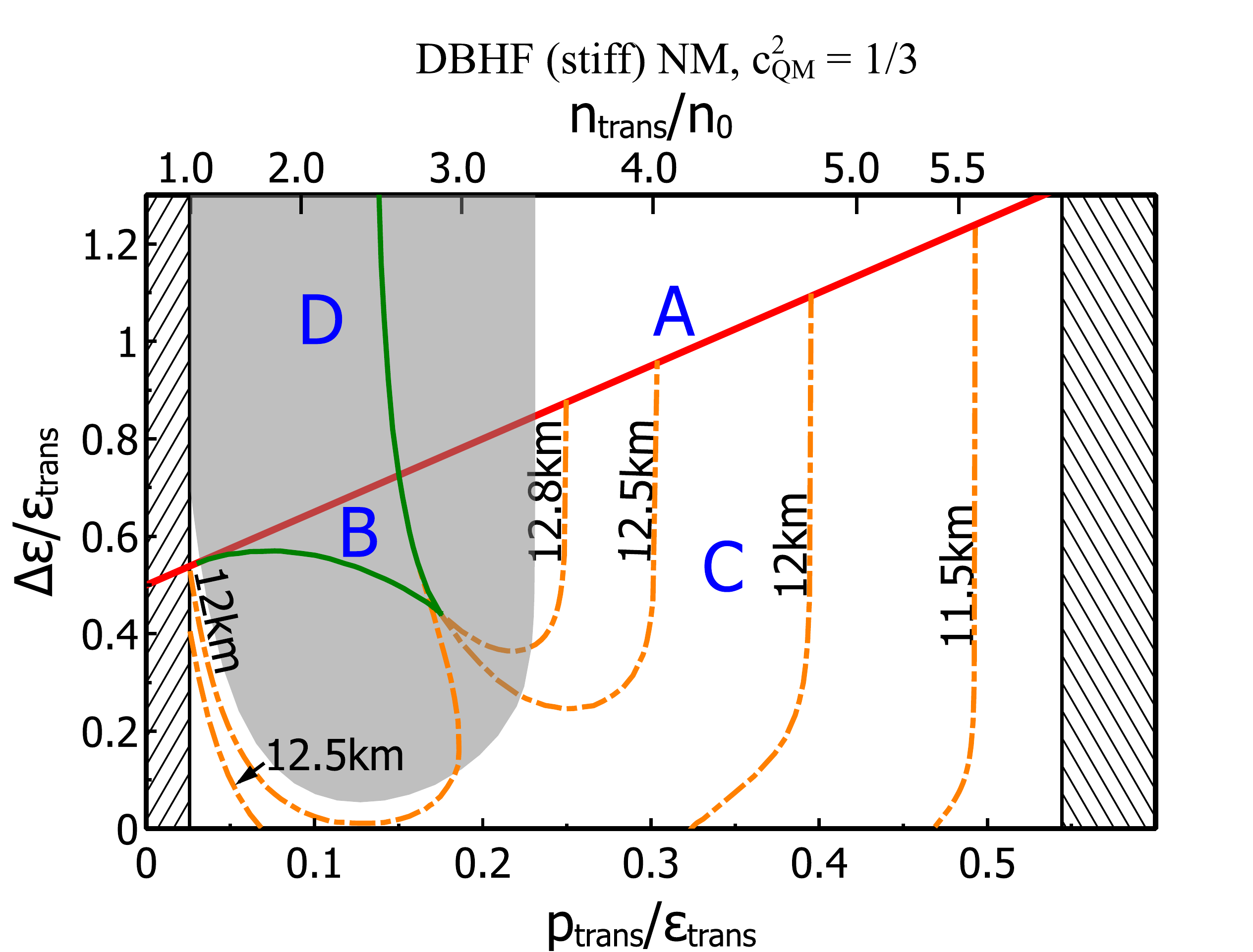}\\[2ex]
}\parbox{0.5\hsize}{
\includegraphics[width=\hsize]{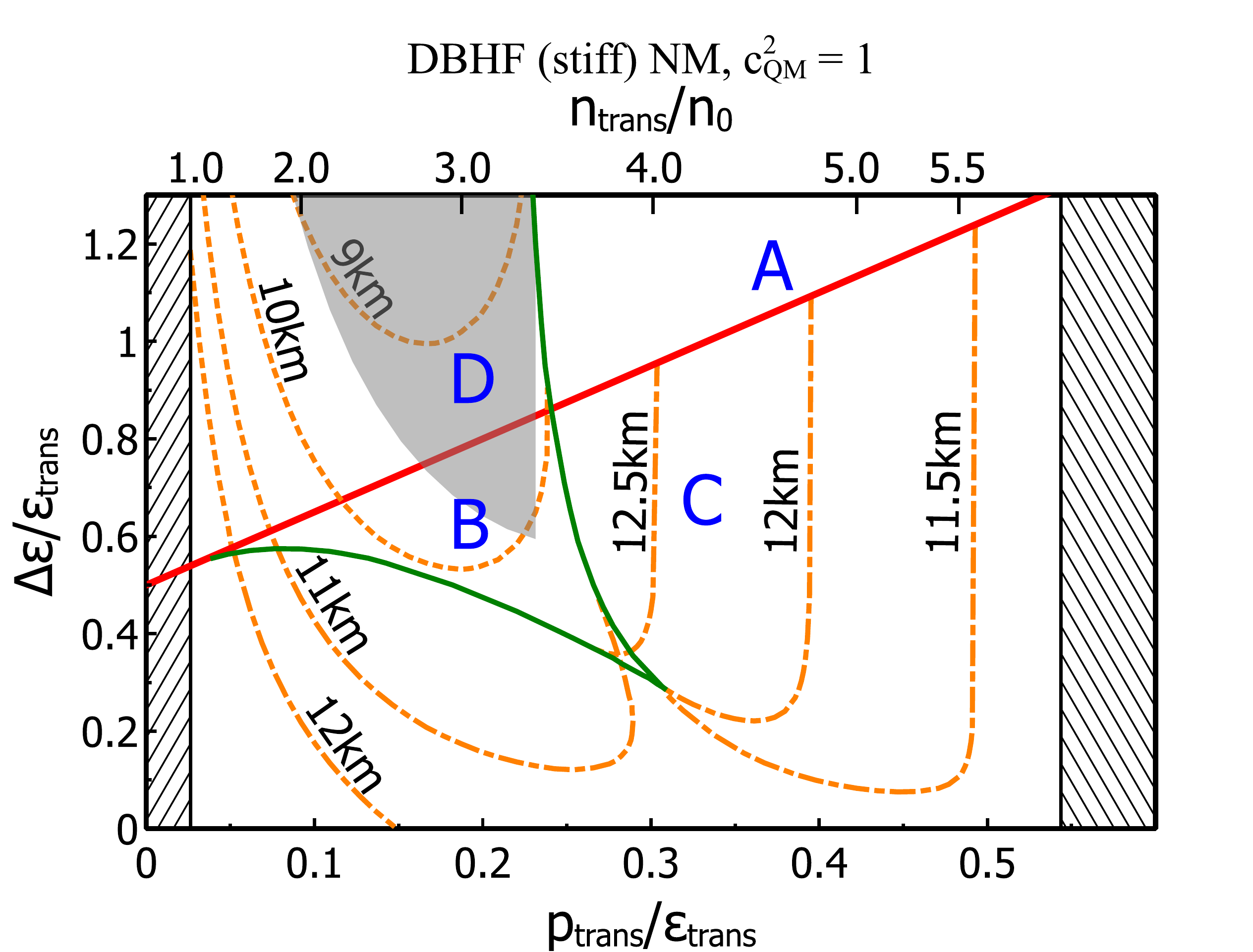}\\[2ex]
}\\[2ex]
\parbox{0.5\hsize}{
\includegraphics[width=\hsize]{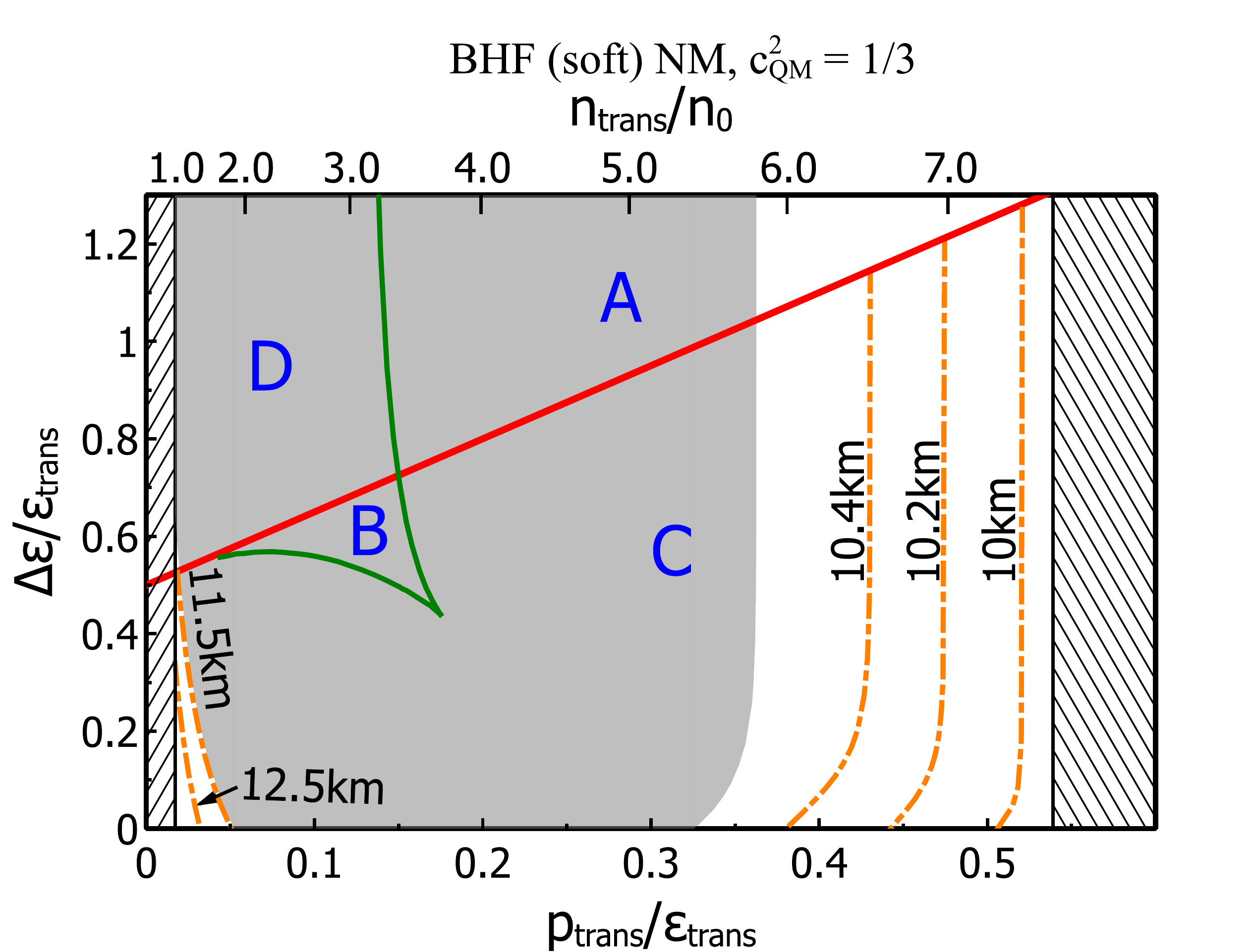}\\[2ex]
}\parbox{0.5\hsize}{
\includegraphics[width=\hsize]{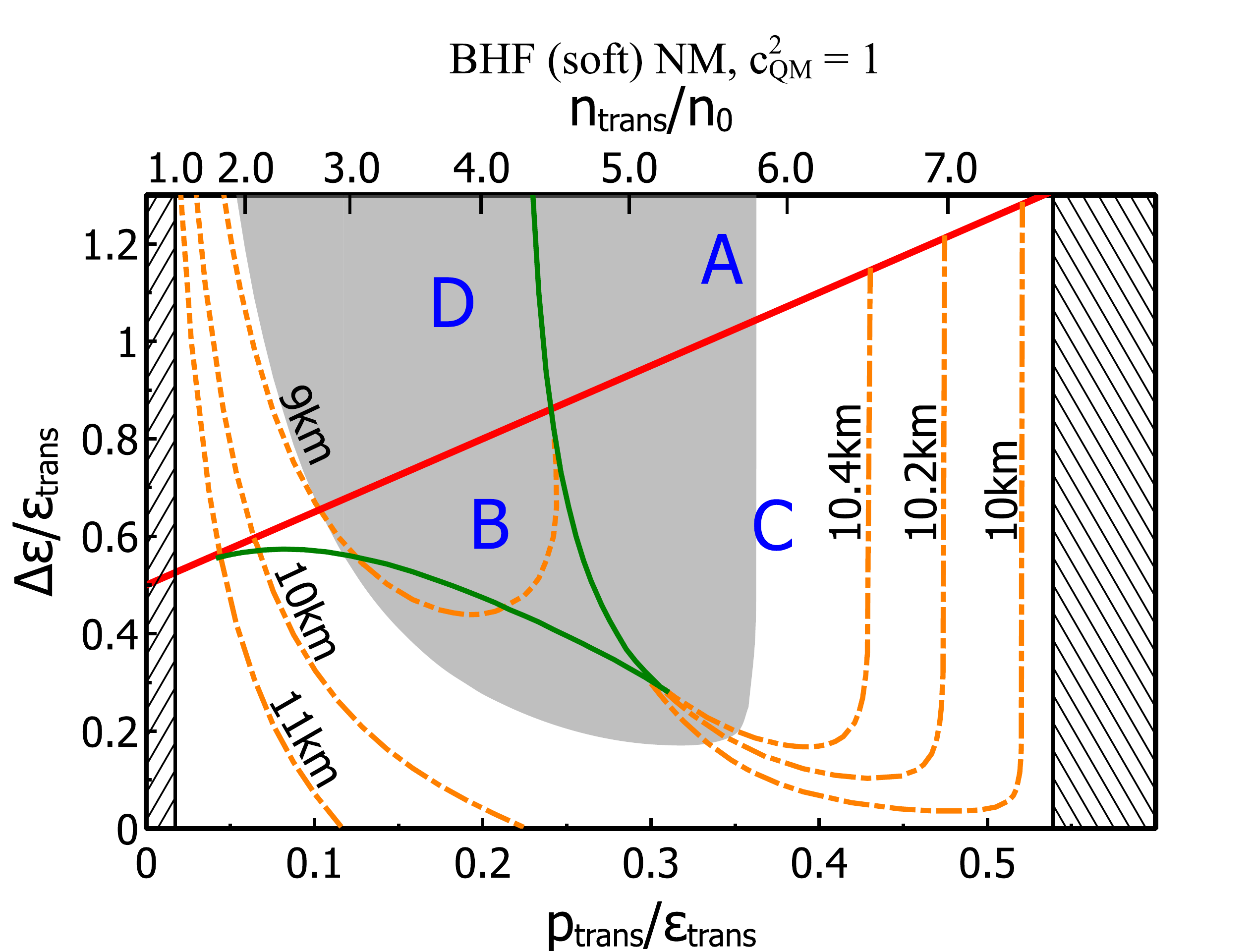}\\[2ex]
}
\caption{(Color online) Contour plots showing the radius of the maximum-mass star as a function of the
CSS parameters. Dashed lines are for the case where this star is on the
disconnected branch; for dot-dashed lines it is on the connected branch.
The grey shaded region where $\Mmax < 1.95\,\Msolar$
is excluded by the measurement of a $2\,\Msolar$ star.
The hatched band at low density (where $\ntrans<n_0$) is
excluded because bulk nuclear matter would be metastable. The
hatched band at high density is excluded because the transition pressure
is above the central pressure of the heaviest stable hadronic star.
For a magnified version of the low-transition-pressure region for $\cQMsq=1/3$,
see Fig.~\ref{fig:CSS-radius-zoom}.
}
\label{fig:CSS-radius-max-mass}
\end{figure*}

In Fig.~\ref{fig:CSS-max-mass}, we show how mass measurements of
neutron stars can be expressed as constraints on the CSS parameters; the upper plots are for DBHF and
the lower plots are for BHF nuclear matter EoS.
Each panel shows dependence on $\ptrans/\etrans$ and $\De\ep/\etrans$
for fixed $\cQMsq$.
The region in which the transition to quark matter would occur below
nuclear saturation density ($\ntrans<n_0$) is excluded (hatched band
at left end) because
in that region bulk nuclear matter would be metastable.
There is also an upper limit on the transition pressure, which is
the central pressure of the heaviest stable nuclear matter star.
This depends on the hadronic EoS that has been assumed.

The contours show the maximum mass of a hybrid star as a function of
the EoS parameters.
The grey shaded region, where $\Mmax < 1.95\,\Msolar$, contains
quark matter EoSs that are
excluded at the one to two $\sigma$ level
by the recent observations of stars of mass $M\approx 2\,\Msolar$
\cite{Antoniadis:2013en} \cite{Demorest:2010bx} (in Ref. \cite{Alford:2015dpa} a slightly higher bound $\Mmax < 2\,\Msolar$ was employed).
For high-density EoSs with $\cQMsq=1$ (right-hand plots), this
region is not too large, and leaves a good range of transition
pressures and energy density discontinuities that are compatible
with the observation.
However, for high-density matter with $\cQMsq=1/3$ (left-hand plots), 
which is the typical value in many models
(see Sec.~\ref{sec:intro}), the $\Mmax\gtrsim 2\,\Msolar$ constraint eliminates
a large region of the CSS parameter space. As one would expect, the stiffer EoS gives rise to heavier
(and larger) stars,
and therefore allows a wider range of CSS parameters to be compatible
with the $2\,\Msolar$ measurement. The region B, where connected and disconnected hybrid star branches
can coexist, is excluded for $\cQMsq\leqslant 1/3$, and even for
larger $\cQMsq$ it is only allowed if the nuclear matter EoS is
sufficiently stiff.

In Fig.~\ref{fig:CSS-max-mass} the dot-dashed (red)
contours are for hybrid stars on a connected branch, while the
dashed (blue) contours are for disconnected branches. When crossing the near-horizontal boundary 
from region C to B the maximum mass
of the connected branch smoothly becomes the maximum mass of the 
disconnected branch, therefore the red contour in the C region
smoothly becomes a blue contour in the B and D regions.
When crossing the near-vertical boundary from region C to B a
new disconnected branch forms, so the connected branch (red dot-dashed) contour 
crosses this boundary smoothly.

If, as predicted by many models, $\cQMsq\lesssim 1/3$,
then the existence of a $2\,\Msolar$ star constrains the other CSS parameters
to two regions of parameter space: in the left-hand region the transition occurs
at a fairly low density $\ntrans \lesssim 2\,n_0$; in the right-hand allowed region the transition pressure is
high, and the connected branch contours are, except at very low $\De\ep$, almost vertical,
corresponding to EoSs that give rise to a very small connected hybrid branch, and the maximum mass on this branch
is very close to the mass of the purely-hadronic matter
star with $p_{\rm cent}=\ptrans$. 
The mass of such a purely-hadronic star is naturally independent
of parameters that only affect the quark matter EoS, such as 
$\De\ep$ and $\cQMsq$, so the contour is vertical. These hybrid stars
have a tiny core of the high density phase and cover a tiny range of masses,
of order $10^{-3}\,\Msolar$ or less (see e.g., Fig.~5 in \cite{Alford:2013aca}), and so would be very rare.

\subsection{Typical radius of hybrid stars}

\begin{figure*}[htb]
\parbox{0.5\hsize}{
\includegraphics[width=\hsize]{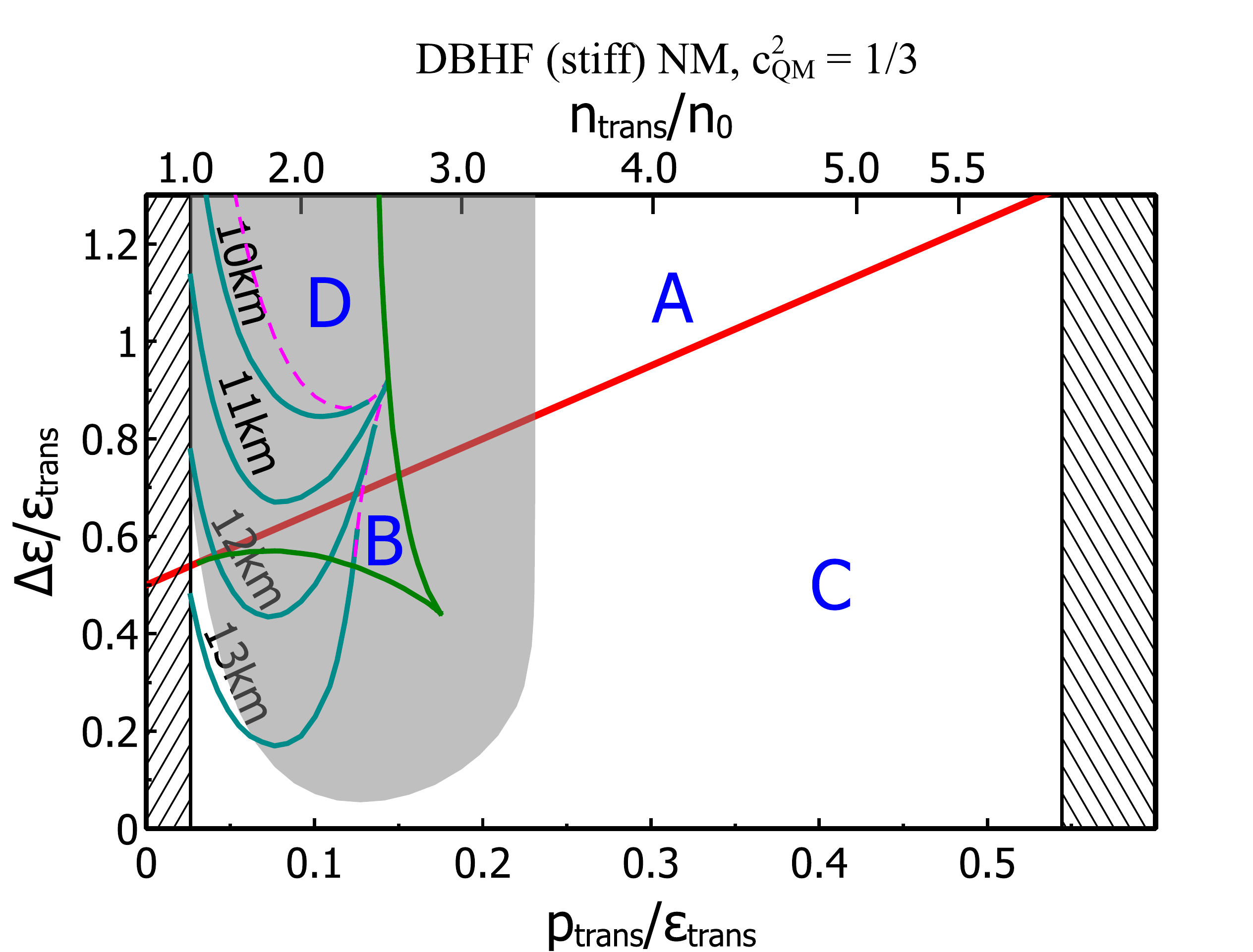}
}\parbox{0.5\hsize}{
\includegraphics[width=\hsize]{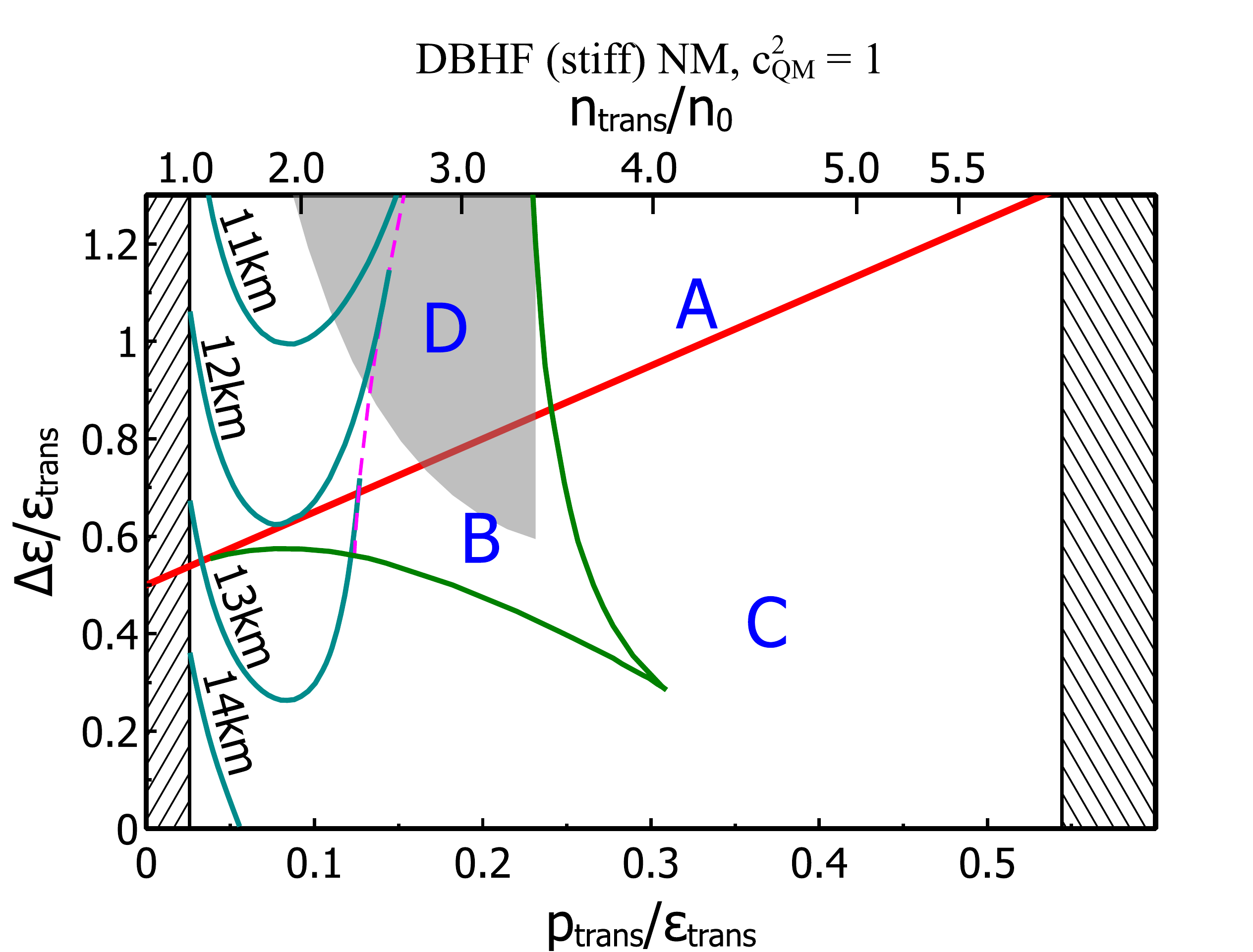}
}\\[2ex]
\parbox{0.5\hsize}{
\includegraphics[width=\hsize]{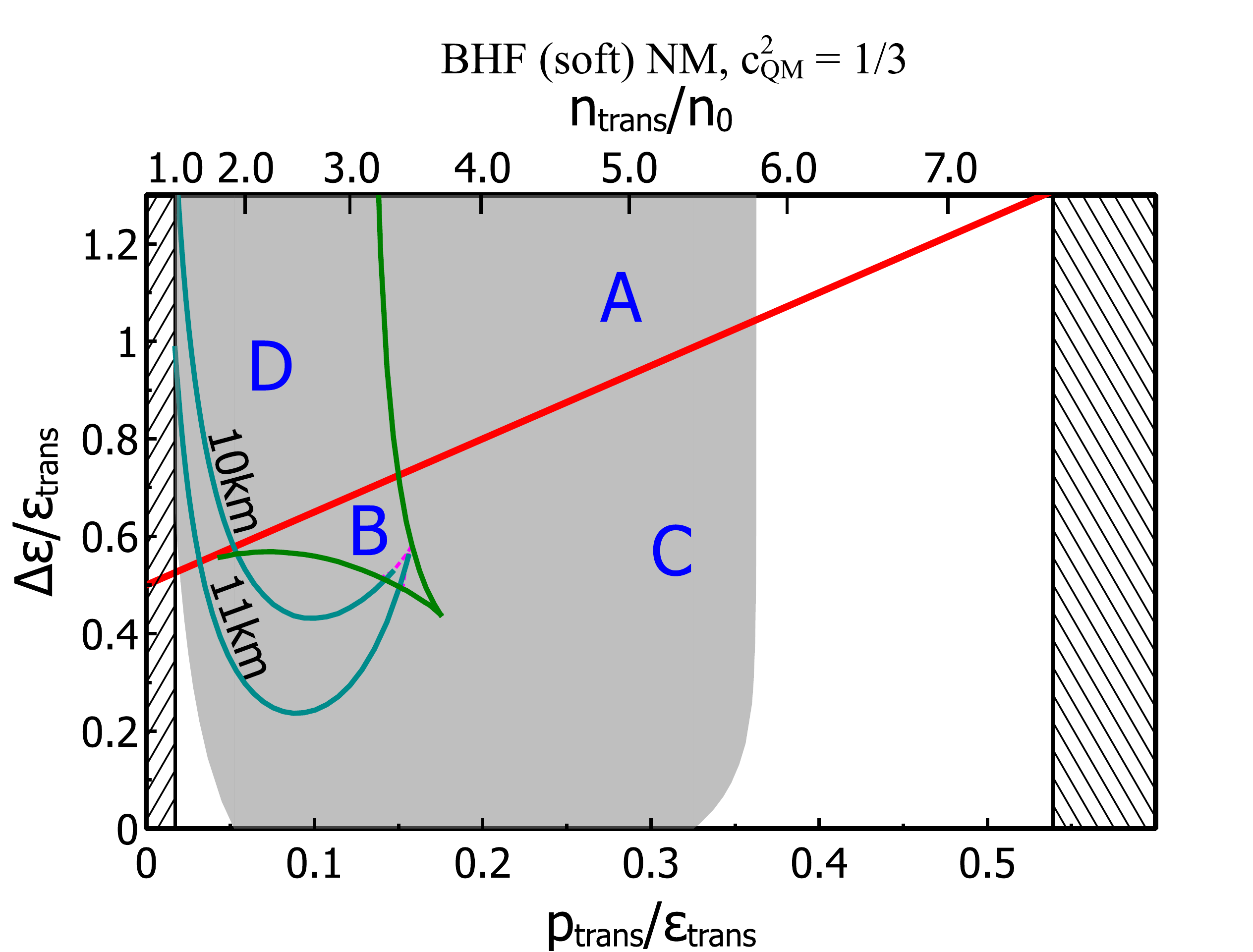}
}\parbox{0.5\hsize}{
\includegraphics[width=\hsize]{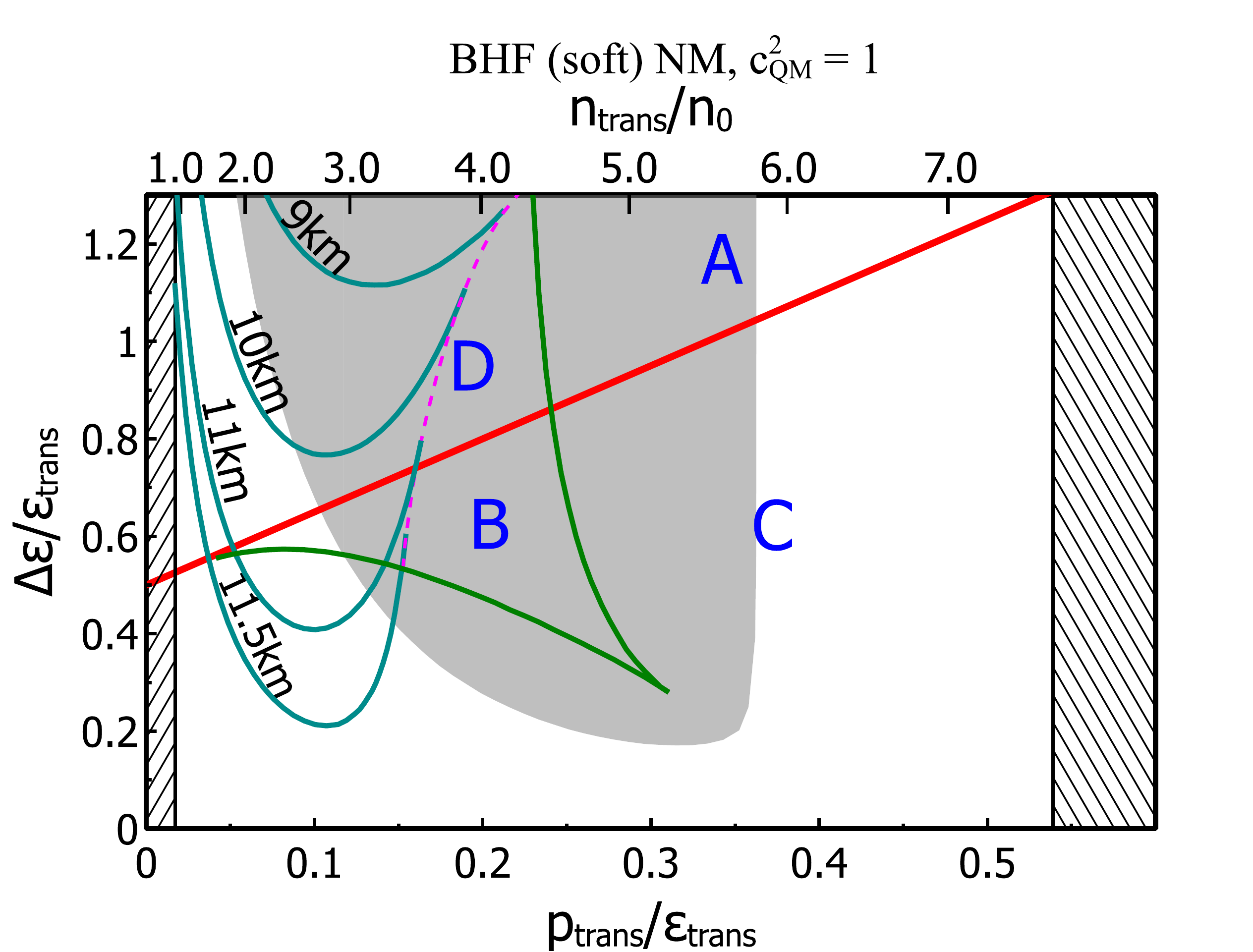}
}
\caption{(Color online) Contour plots similar to Fig.~\ref{fig:CSS-radius-max-mass}
showing the radius of a hybrid star of mass $M=1.4\,\Msolar$
as a function of the CSS parameters. Such stars only exist in a limited
region of the space of EoSs (delimited by dashed (magenta) lines).
Outside that region the only $1.4\,\Msolar$ star is a hadronic star
with radius $11.8$\,km (BHF) or $13.4$\,km (DBHF) 
(see Table \ref{tab:EoS}).
The grey shaded region where $\Mmax < 1.95\,\Msolar$
is excluded by the measurement of a $2\,\Msolar$ star.
For a magnified version of the low-transition-pressure region for $\cQMsq=1/3$,
see Fig.~\ref{fig:CSS-radius-zoom}.
}
\label{fig:CSS-radius}
\end{figure*}

\begin{figure*}[htb]
\parbox{0.5\hsize}{
\includegraphics[width=\hsize]{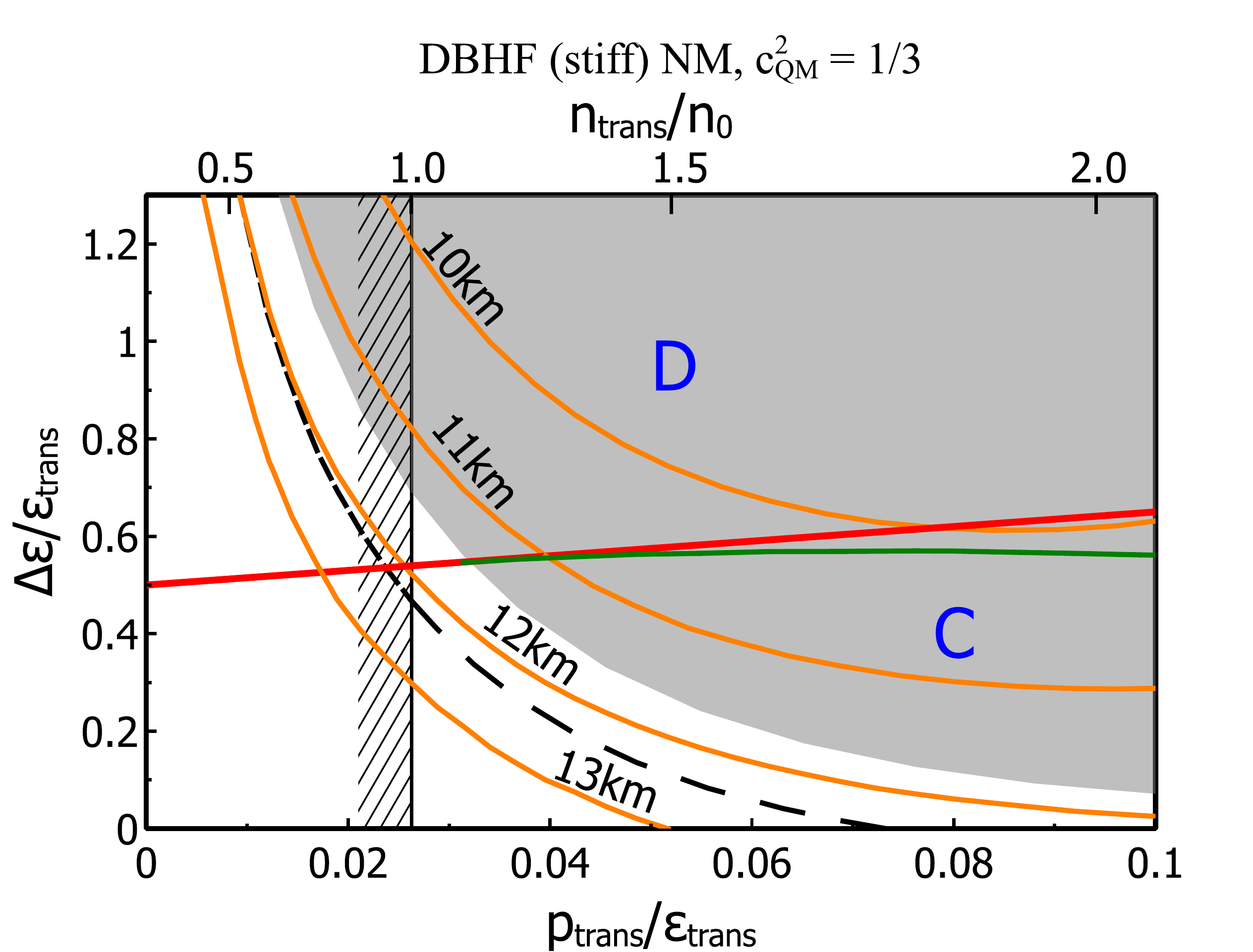}
}\parbox{0.5\hsize}{
\includegraphics[width=\hsize]{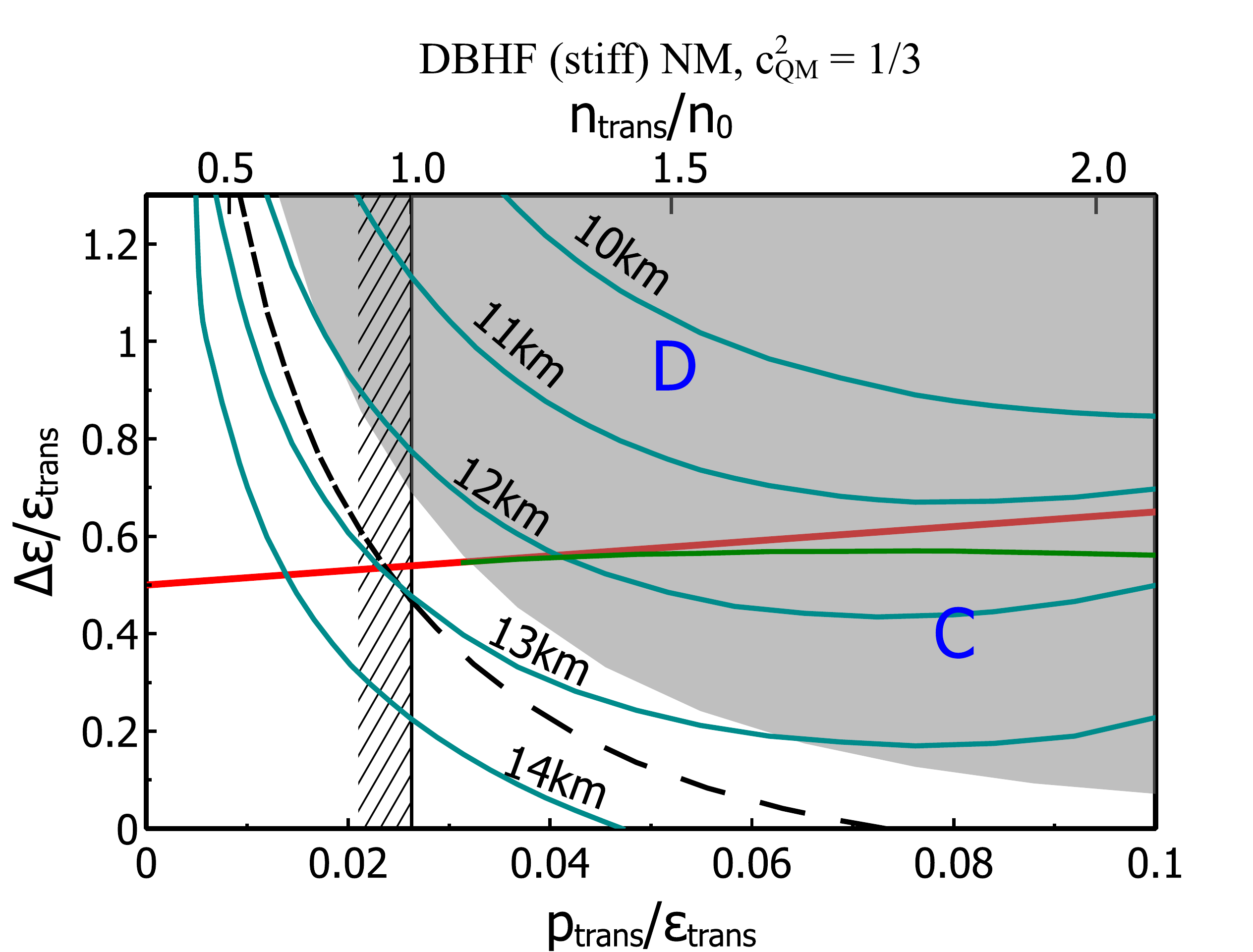}
}\\[2ex]
\parbox{0.5\hsize}{\includegraphics[width=\hsize]{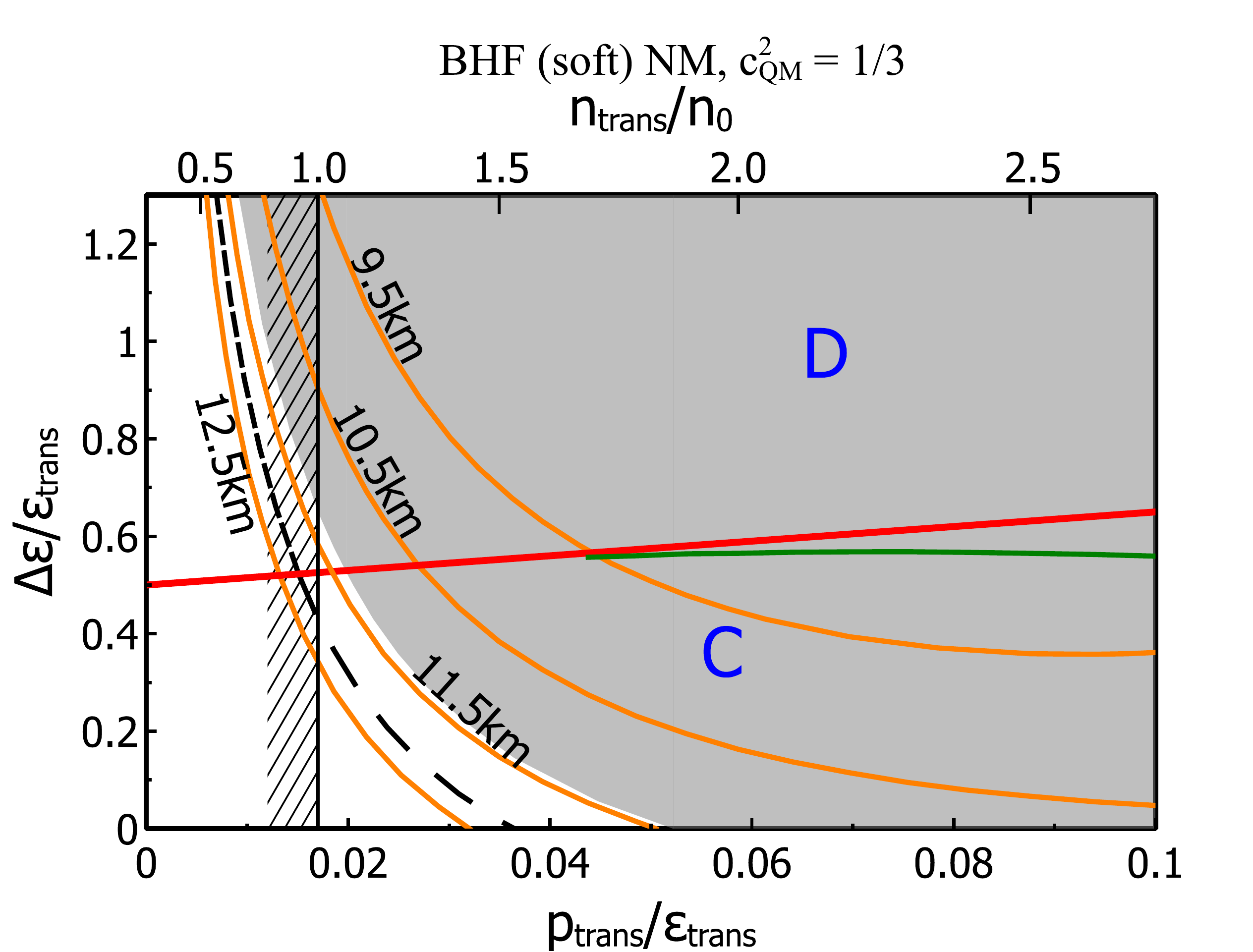}
}\parbox{0.5\hsize}{
\includegraphics[width=\hsize]{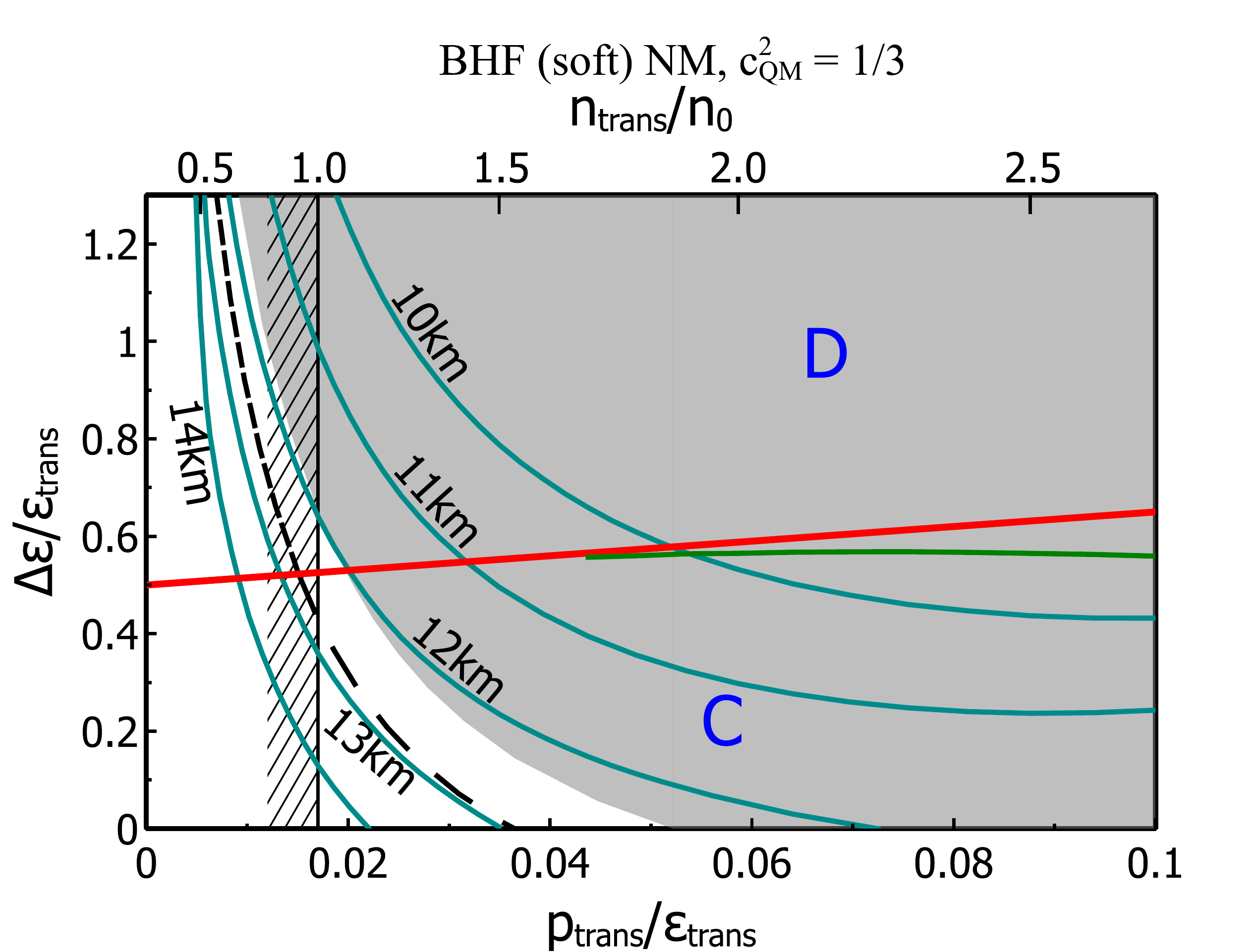}
}
\caption{(Color online) Magnified version of the $\cQMsq=1/3$ plots
in Figs.~\ref{fig:CSS-radius-max-mass} and \ref{fig:CSS-radius}.
In the two left panels, the contours are for the radius of the maximum mass star, which is typically the smallest star
for a given EoS. In the two right panels, the contours are for
$R_{1.4}$, the radius of a $1.4\,\Msolar$ star. 
The region under and to the left of the hatched bar
is probably unphysical because $\ntrans < n_0$, and it was 
excluded (hatched band) in earlier figures.
The grey shaded region where $\Mmax < 1.95\,\Msolar$
is excluded by the measurement of a $2\,\Msolar$ star.
The dashed line shows how that region would grow if a
$2.1\,\Msolar$ star were observed.
}
\label{fig:CSS-radius-zoom}
\end{figure*}

\subsubsection{Minimum radius}
In Fig.~\ref{fig:CSS-radius-max-mass} we show contour plots of the 
radius of the maximum-mass star (on either a connected or disconnected hybrid branch) as a function of the CSS quark matter
EoS parameters. Since the smallest hybrid star is typically the heaviest
one, this allows us to infer the smallest radius that arises from
a given EoS.

\begin{figure*}[htb]
\parbox{0.5\hsize}{
\includegraphics[width=\hsize]{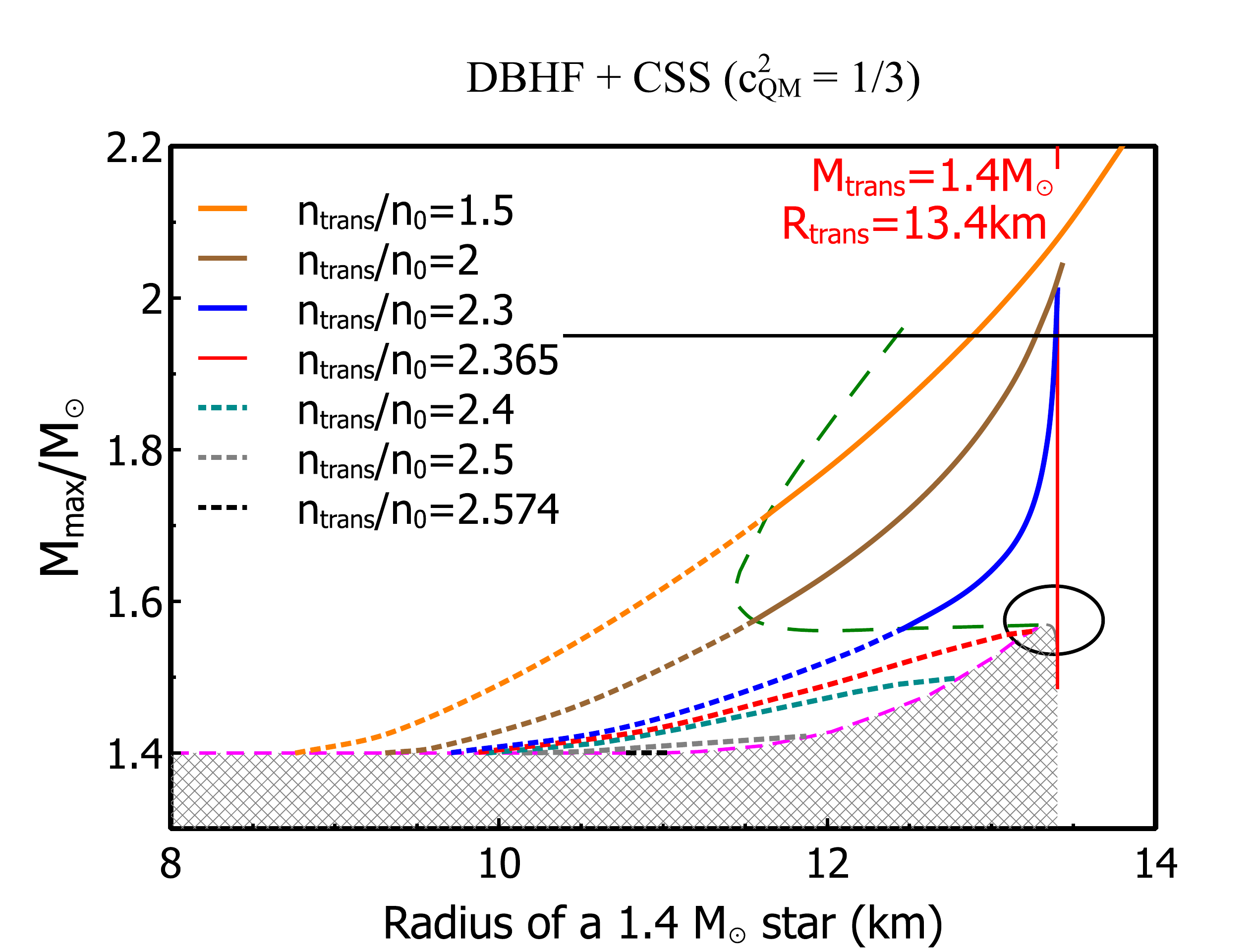}
}\parbox{0.5\hsize}{
\includegraphics[width=\hsize]{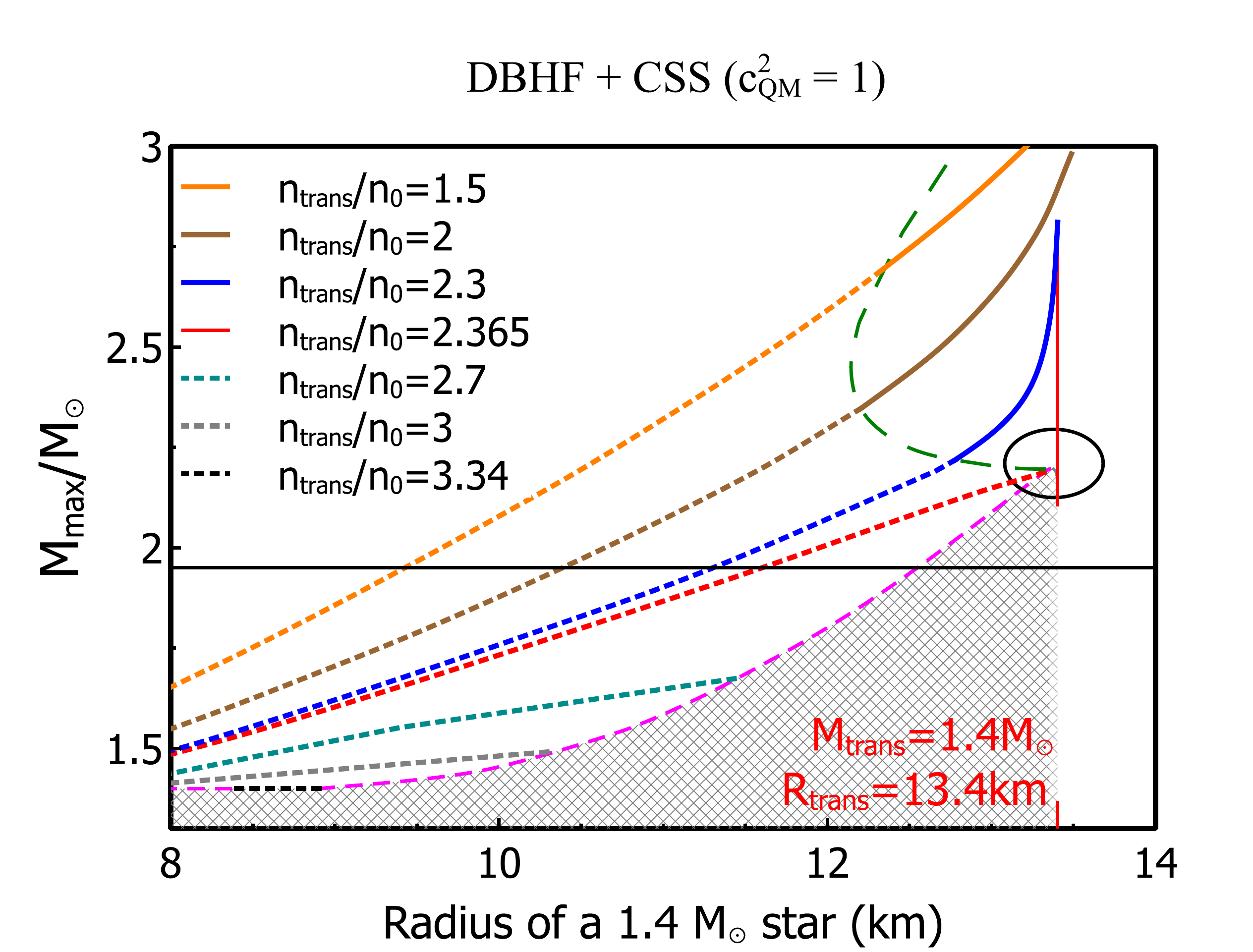}
}\\[2ex]
\parbox{0.5\hsize}{
\includegraphics[width=\hsize]{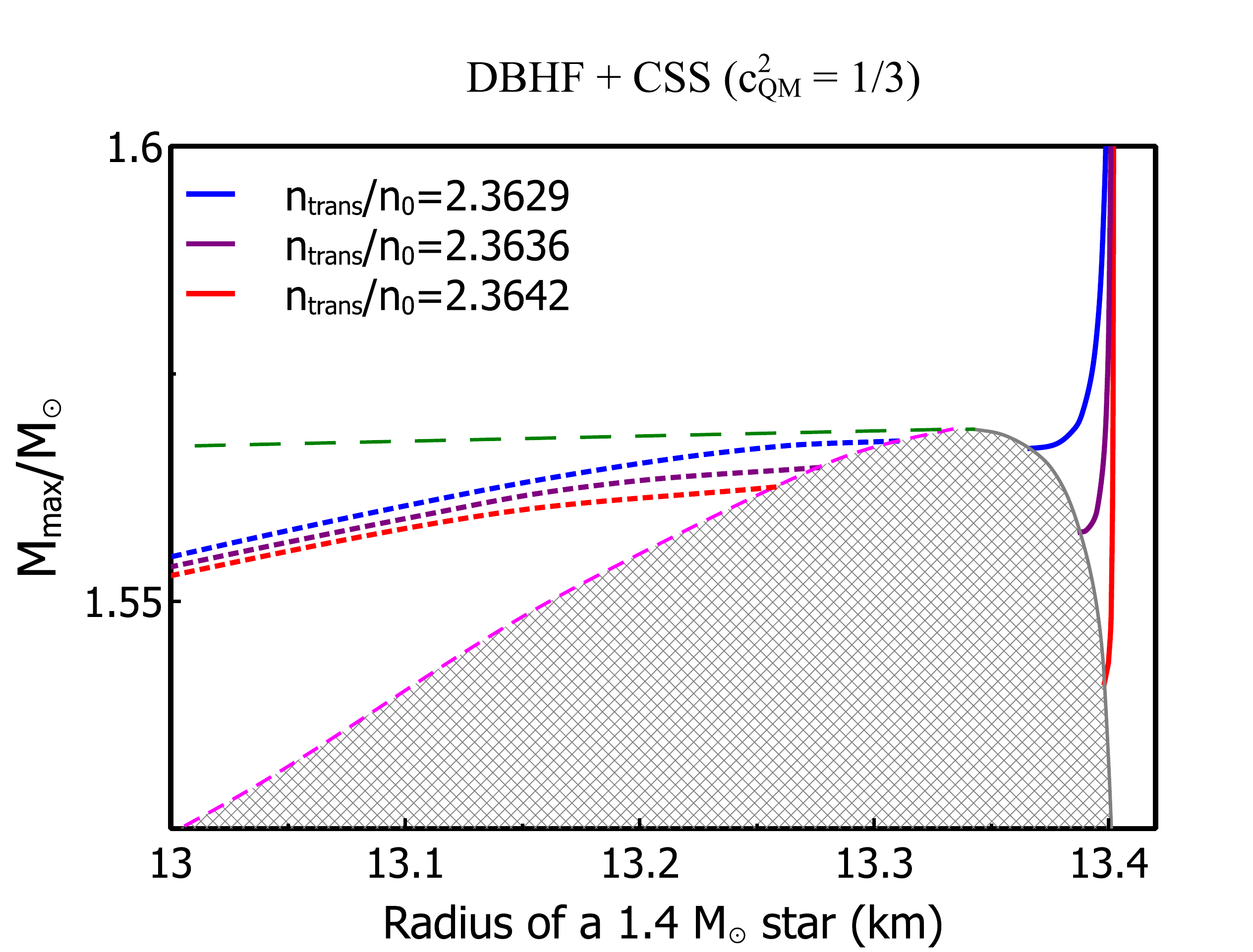}
}\parbox{0.5\hsize}{
\includegraphics[width=\hsize]{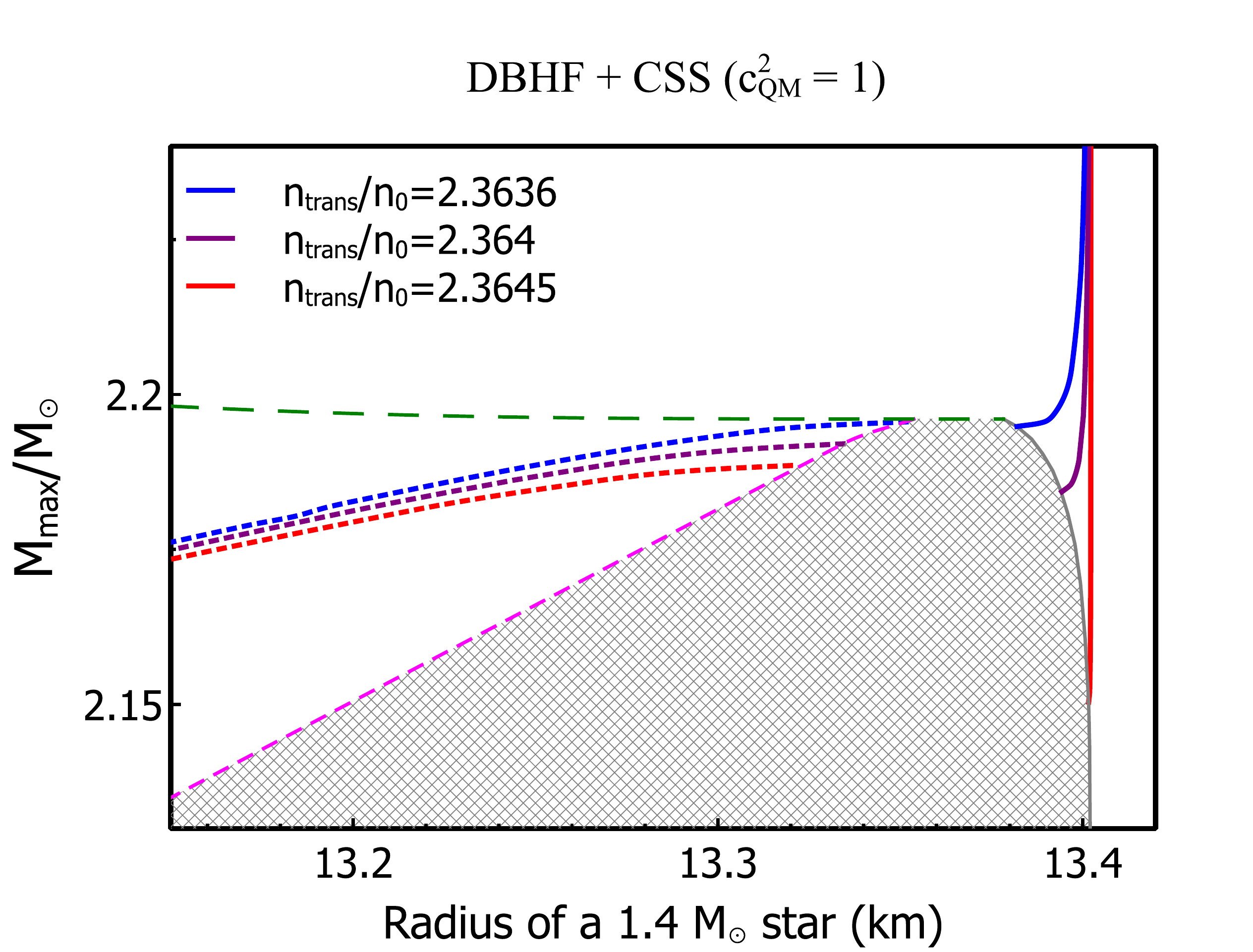}
}
\caption{(Color online) Contour plots showing how the maximum hybrid star mass and the radius of a $1.4\,\Msolar$ star vary when two of the CSS parameters, the transition density and energy density discontinuity, are varied. Shaded regions indicate where no $1.4\,\Msolar$ hybrid star exists. In the upper panels, EoSs below the
horizontal lines (i.e. those with $\Mmax<1.95\,\Msolar$)
are ruled out by observation, and the vertical (red) lines mark the radius of a $1.4\,\Msolar$ purely-hadronic star. Lower panels are zoomed in on the encircled region in the upper panels, where EoSs can give a $1.4\,\Msolar$ hybrid star on both connected branch (solid curves on the right) and disconnected branch (dashed curves on the left).
}
\label{fig:CSS-max-mass-r-1.4}
\end{figure*}

The smallest stars, with radii as small as 9\,km, occur when
the high-density phase has the largest possible speed of sound
$\cQMsq=1$. They are disconnected branch stars arising from EoSs having
a low transition pressure ($\ntrans \lesssim 2\,n_0$) 
with a fairly large energy density discontinuity 
($\De \ep/\etrans \gtrsim 1$). The contours in the
high-transition-pressure region are almost vertical because the hybrid branch
is a very short extension to the nuclear mass-radius relation, and its radius decreases with $\ptrans$ in this region, because it decreases with central pressure.

For $\cQMsq=1/3$, the allowed low-transition-pressure region is
disconnected from the high-transition pressure region and is
so small that it is hard to see on this plot. By magnifying it (left-hand
plots of Fig.~\ref{fig:CSS-radius-zoom}) we see that in this region
the radius contours closely track the border of the allowed region
(the $M_{\rm max}=1.95\,\Msolar$ line)
so we can say that the radius must be greater than 11.25\,km.
For a stiff hadronic EoS this
minimum is raised to 11.4\,km. These values are comparable to
the minimum
radius of about 11.8\,km found in Ref.~\cite{Bedaque:2014sqa}, which
explored a larger set of hadronic EoSs but did not
explore the full CSS parameter space for the high-density EoS.
If a star with radius smaller than this minimum value were to be observed,
we would have to conclude that either the transition
occurs outside the low-density region or that $\cQMsq$ is greater than 1/3.
In the magnified figure we also show how the excluded region would
grow if a $2.1\,\Msolar$ star were to be observed (long-dashed line
for connected branch stars and short-dashed line for disconnected branch stars).
This would increase the minimum radius to about 12.1\,km for the soft hadronic EoS and 12.2\,km for the stiff hadronic EoS.

\subsubsection{Radius of a $1.4\,\Msolar$ star}
\label{sec:R1.4}

In Fig.~\ref{fig:CSS-radius} we show contours (the U-shaped lines) of
typical radius of a hybrid star, defined as
$R_{1.4}$, the radius of a star of mass $1.4\,\Msolar$,
as a function of the CSS parameters. 
The contours only fill the part of the CSS parameter space
where there are hybrid stars with that mass.
The dashed (magenta) lines delimit that region
which extends only up to moderate transition pressure.

The overall behavior is that, at fixed $\De\ep/\etrans$, the typical radius
is large when the transition density is at its lowest. As the transition
density rises the radius of a $1.4\,\Msolar$ star decreases at first, but then 
increases again. This is related to the
previously noted fact \cite{Yudin:2014mla}
that when one fixes the speed of sound of quark
matter and increases the bag constant 
(which increases $\ptrans/\etrans$ and also varies
$\De\ep/\etrans$ in a correlated way)
the resultant family of mass-radius curves
all pass through the same small region
in the $M$-$R$ plane: the $M(R)$ curves ``rotate'' counter-clockwise around
this hub (see Fig.\,2 of Ref.~\cite{Yudin:2014mla}).
In our case we are varying $\ptrans/\etrans$ at fixed $\De\ep/\etrans$,
so the hub itself also moves. At low transition density the hub is
below $1.4\,\Msolar$, so $R_{1.4}$ decreases with $\ptrans/\etrans$.
At high transition density the hub is at a mass above $1.4\,\Msolar$ 
so $R_{1.4}$ will increase with $\ptrans/\etrans$.

The smallest stars occur for $\cQMsq=1$ (right-hand plots), 
where $R_{1.4}\gtrsim 9.5\,{\rm km}$
at large values of the energy density discontinuity, and the radius
rises as the discontinuity is decreased. This is consistent
with the absolute lower bound of about 
8.5\,km \cite{Lattimer:2012nd} for the
maximally compact $\cQMsq=1$ star obeying $\Mmax> 2\,\Msolar$.

For $\cQMsq=1/3$ the allowed region at low transition pressure is small,
so in the right panels of Fig.~\ref{fig:CSS-radius-zoom} we show a magnification
of this region. We see that in the allowed ($\Mmax>1.95\,\Msolar$ and $\ntrans>n_0$) region 
there is a minimum radius $12\,{\rm km}$ for the BHF (soft) hadronic EoS,
and about $12.25\,{\rm km}$ for the DBHF (stiff) hadronic EoS. This
minimum is attained at the lowest possible transition density,
$\ntrans\approx n_0$. As the transition density rises to values
around $2\,n_0$, the minimum radius rises to 12.4\,km (BHF)
or 13.3\,km (DBHF). This is comparable to the minimum
radius of about 13\,km found in Ref.~\cite{Bedaque:2014sqa}, which
explored a wider range of hadronic EoSs but assumed $\ntrans=2\,n_0$.
These results are consistent with
the lower bound on $R_{1.4}$ for $\cQMsq=1/3$ of about 
11\,km established in Ref.~\cite{Lattimer:2012nd} (Fig.~5)
using the EoS that yields maximally compact stars
(corresponding to CSS with $\ptrans=0$ and $\cQMsq=1/3$) 
obeying  $\Mmax>2\,\Msolar$.
If a $1.4\,\Msolar$ star were observed to have radius below the minimum value, one would
have to conclude that it could only arise from a CSS-type EoS
if that EoS had $\cQMsq > 1/3$.
The dashed line shows how the excluded region would grow if a
star of mass $2.1\,\Msolar$ were to be observed. 
This would increase the minimum
radius to about 12.7\,km (BHF) or 13\,km (DBHF). 

\subsection{Maximum mass vs.~typical radius}

Following Ref.~\cite{Lattimer}, in Fig.~\ref{fig:CSS-max-mass-r-1.4} we
characterize each CSS EoS by the
maximum mass $\Mmax$ and the radius $R_{1.4}$ of a $1.4\,\Msolar$ star
on its mass-radius relation. We
can then see which areas of the $R_{1.4}$-$\Mmax$ plane can be populated by
hybrid stars arising from typical nuclear EoS combined with the CSS family of
quark matter EoSs.  Along each curve the transition density is fixed and we
vary $\De\ep/\etrans$.

Each of the monotonic curves in Fig.~\ref{fig:CSS-max-mass-r-1.4}
represents a family of CSS EoSs with fixed transition density, and varying
$\De\ep/\etrans$. As we increase $\De\ep/\etrans$ (from right to left along the curve), the maximum mass on a connected branch (solid curve) decreases until it smoothly becomes the maximum mass on a disconnected branch (dashed curve) at the long-dashed (green) line, which corresponds to the nearly-horizontal phase boundary
between region B and region C in the phase diagram
Fig.~\ref{fig:phase-diag-HLPS}. For higher transition densities the whole
curve is dashed, because in that case any mass-radius curve that includes a
$1.4\,\Msolar$ hybrid star will always have its heaviest star on the
disconnected branch. 

The thin vertical (red) line is where 
the transition pressure has risen so high that it is equal to
the central pressure of a nuclear matter star with
a mass of $1.4\,\Msolar$ (at $\ntrans=2.365 \,n_0$ for DBHF nuclear EoS). Some of the mass contours extend to the right of the vertical (red) line, where a 1.4 $\Msolar$ hybrid star can have larger radius than the purely-hadronic star. This corresponds to phase transitions at low densities ($\ntrans \lesssim 2 \,n_0$) with a small energy density jump ($\De\ep/\etrans \lesssim 0.1$), when at central pressures above transition the hybrid star M-R relation starts to acquire a similar behavior to that of strange stars, i.e. the radius increases with the mass.
At higher transition pressures, although on the connected branch hybrid stars are always above $1.4\,\Msolar$, a $1.4\,\Msolar$ hybrid star on the disconnected branch can still exist (see contours in upper panels of Fig.~\ref{fig:CSS-max-mass-r-1.4}, where the whole
curve is dashed for $\ntrans>2.365 \,n_0$).

On the lower panels we show contours for EoSs that can give hybrid stars on both connected branch (solid curves on the right) and disconnected branch (dashed curves on the left), at transition densities very close to but below the central density of a nuclear matter star with
a mass of $1.4\,\Msolar$ ($\ntrans<2.365 \,n_0$). From these plots we can see that with the same maximum mass, radii for $1.4\,\Msolar$ hybrid stars on connected/disconn-
ected branches can differ by from $0.1$ to $0.4 \rm\,km$.

Each plot in Fig.~\ref{fig:CSS-max-mass-r-1.4} contains a
black horizontal line at $\Mmax=1.95\,\Msolar$, so all EoSs that lie below
that line are observationally ruled out.
If $\cQMsq \lesssim 1/3$ (see left-hand plots), the range of radii for $1.4\,\Msolar$ hybrid stars is limited to the vicinity of the 
purely-hadronic star with mass $1.4\,\Msolar$. This is another way
of illustrating the lower limit $R>12.25\,\km$ found in Sec.~\ref{sec:R1.4}
for the DBHF nuclear EoS.
If $\cQMsq = 1$ (right-hand plots) the range of possible radii
is somewhat bigger, $R>9.5\, \km$, compatible with the radius value  tracking along the $\Mmax=1.95\,\Msolar$ boundary in Fig.~\ref{fig:CSS-radius}, upper-right panel.

\section{Phase conversion dissipation in multicomponent compact stars}
\label{sec:phase-conversion}

\begin{figure}[htb]
\includegraphics[width=\hsize]{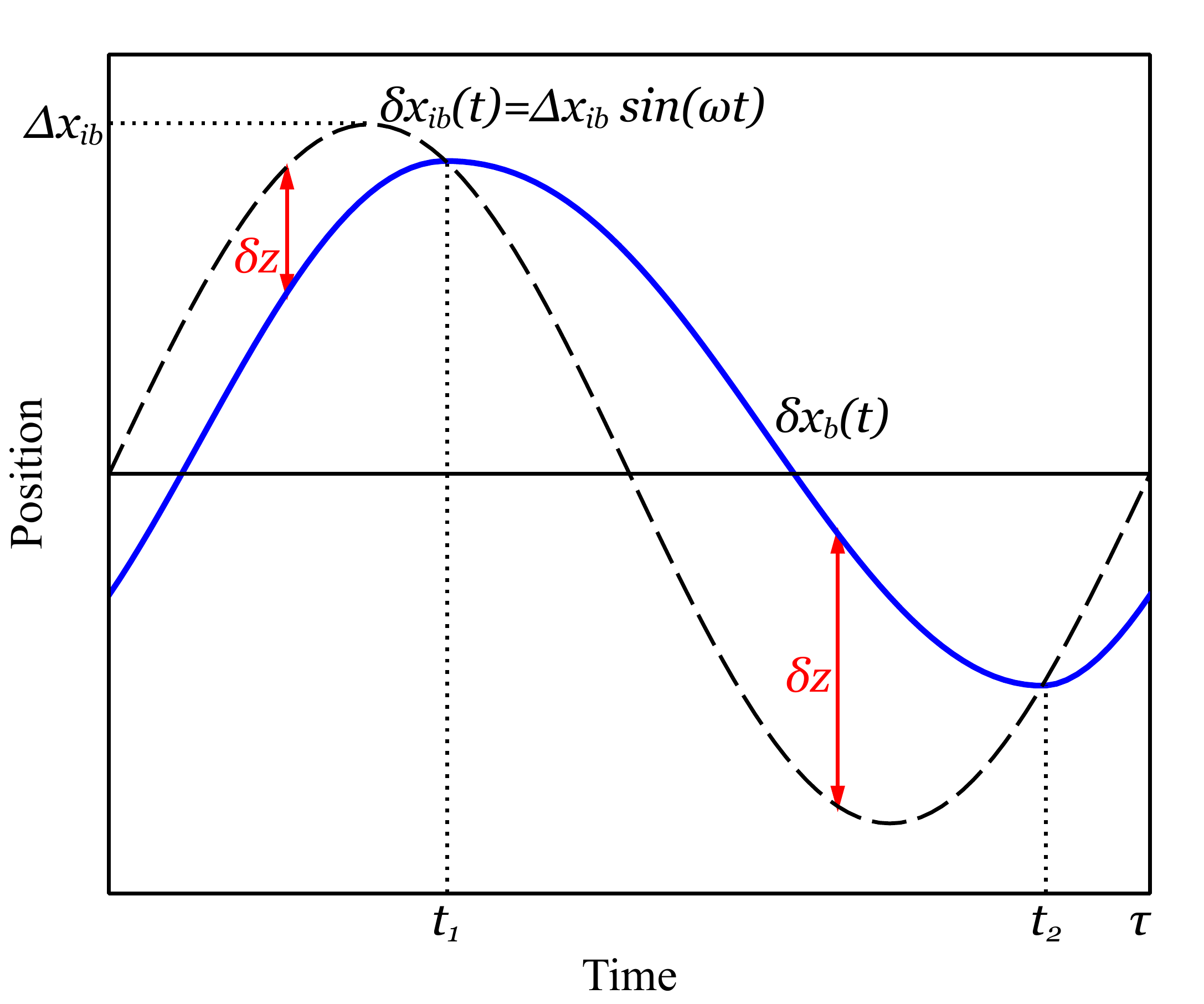}
\caption{(Color online) Diagram showing how the ideal boundary position (dashed line) and the real boundary position (solid blue line) vary in time. The ideal boundary is where the phase boundary would be if the phase conversion process equilibrated instantaneously, and it is determined by the instantaneous external pressure. The real boundary is always ``chasing'' the ideal
boundary, with velocity given by Eq.~\eqn{eqn:dx_t_qm} and \eqn{eqn:dx_t_nm}
where $\de z(t)$ is its distance from the ideal boundary. The real boundary coincides with the ideal boundary twice per cycle, at $t=t_1$ and $t=t_2$.}
\label{fig:x_ideal_real}
\end{figure}

As noted in Sec.~\ref{sec:intro}, it is very important to understand
the dissipation mechanisms that suppress oscillations in neutron stars.
One example is the damping of r-modes, which controls the spin down of
neutron stars. Observational data indicates that the conventional
damping mechanisms (bulk and shear viscosity) are insufficient to
explain the existence of millisecond pulsars with the observed
temperatures \cite{Ho:2011tt,Alford:2013pma}.

Here we describe a dissipation mechanism that could account for the
observations by saturating the growth of r-modes at a very low amplitude
\cite{Alford:2014jha}.  The mechanism is based on the movement of the
core-mantle interface of a hybrid star in response to the pressure
oscillations in the star. This requires conversion of hadronic matter into
quark matter and vice versa.  If the finite rate of this conversion produces a
phase lag between the pressure oscillation and the position of the interface,
energy will be dissipated in each cycle. We study the resultant damping and we
find that this mechanism is powerful enough to saturate the r-mode at very low
saturation amplitude, of order $10^{-10}$, and is therefore likely to be the
dominant r-mode saturation mechanism in hybrid stars with a sharp interface.

\subsection{Hadron-quark conversion in a hybrid star}
\label{sec:hadron-quark-conversion}

In a one-dimensional, simplified model describing two phases in a cylinder separated by a sharp boundary moving in response to external oscillation in the context of Newtonian gravity, the expression of the corresponding energy dissipation in one cycle of oscillation as a function of the boundary velocity is given as \cite{Alford:2014jha}
\beq
\ba{rcl}
W &=& \dsp S\left(\frac{n_{\rm H}}{n_{\rm L}}-1\right) \left( \int_0^\tau \overline{p}\frac{\mathrm{d}\delta x_{\rm b}(t)}{\mathrm{d}t}\,\mathrm{d}t \right. \\[3ex]
&&\left.\qquad \qquad\quad \dsp + \int_0^\tau \Delta{p_{\rm L}}\sin(\omega t)\frac{\mathrm{d}\delta x_{\rm b}(t)}{\mathrm{d}t}\,\mathrm{d}t \right).
\ea
\label{eq:W1_W2} 
\eeq
assuming that the pressure in the low-density phase oscillates harmonically with amplitude $\Delta p_{\rm L}$ and frequency $\omega$ around its equilibrium value $\overline p$, and $\overline{x}_{\rm b}$ is the equilibrium position of the boundary (the center of the star is at $x=0$). Here we look at the case where the phases are quark matter and hadronic matter and are interested in situations where phase conversion dissipation
becomes important in r-mode oscillations when their amplitude is still
fairly low, assuming
$\delta p \ll \overline{p}$ in the region near the boundary. 
Therefore, we only need the EoS in a narrow pressure range
around the critical pressure at transition. 

We use the calculational techniques developed by Olinto
\cite{Olinto:1986je} to study the movement of the phase boundary in the
strange matter hypothesis scenario, but we are interested in conversion of
quark matter to nuclear matter as well as nuclear matter to quark matter,
since both processes occur as our burning front moves inwards and outwards
periodically in response to an oscillation in the pressure.
The velocity of the boundary in the $\rm NM\to QM$ half cycle 
($t<t_1$ and $t>t_2$ in Fig.~\ref{fig:x_ideal_real}) is determined by \cite{Alford:2014jha}
\beq
\frac{\mathrm {d} \delta x_{\rm b}}{\mathrm {d}t}
\simeq \frac{1}{a_{\rm N}}\sqrt{\frac{D_{\rm Q}}{2\tau_{\rm Q}}}\sqrt{\bigl[\de z / \ell_{\rm Q} \bigr]^2+2\eta_{\rm Q}\de z / \ell_{\rm Q}},
\label{eqn:dx_t_qm} 
\eeq
where $\de z \equiv |\delta x_{\rm ib}-\delta x_{\rm b}|$
is how far the boundary is from its equilibrium 
position at the current pressure (see Fig.~\ref{fig:x_ideal_real}), and $\ell_{\rm Q}$ characterizes its typical length
\beq
\ell_{\rm Q}= \frac{(n_{\rm Q}/\chi_{\rm K}^{\rm Q})n_{\rm Q}}{2(\gamma-1)g_{\rm b}\ep_{\rm crit}^{\rm N}}.
\label{eqn:a_tilde_qm}
\eeq
In Eq.~\eqn{eqn:dx_t_qm}, $a_{\rm N}$ characterizes the strangeness fraction, $D_{\rm Q}$ is the diffusion constant for flavor, $\tau_{\rm Q}$ is the
time scale of nonleptonic flavor-changing interactions, $\eta_{\rm Q}$ gives the ratio of subthermal to suprathermal rates; in Eq.~\eqn{eqn:a_tilde_qm}, $n_{\rm Q}$ is the baryon number density in equilibrated
strange quark matter, $\chi_{\rm K}^{\rm Q}\equiv\partial n_{\rm K}/\partial \mu_{\rm K}$ is the susceptibility with respect
to K-ness evaluated at equilibrium, $\gamma\equiv n_{\rm Q}/n_{\rm N}$ where $n_{\rm N}$ is the baryon number density in 
nuclear matter, $g_{\rm b}$ is the effective gravitational acceleration and $\ep_{\rm crit}^{\rm N}$ is the energy density at the phase boundary on nuclear matter side (see Sec~IV in \cite{Alford:2014jha} for detailed discussions). 

The boundary velocity in the $\rm QM \to NM$ half cycle ($t_1<t<t_2$ in Fig.~\ref{fig:x_ideal_real}) is
\beq
\frac{\mathrm {d} \delta x_{\rm b}}{\mathrm {d}t}
\simeq \frac{1}{b_{\rm Q}}\sqrt{\frac{D_{\rm N}}{2\tau_{\rm N}}}\sqrt{\bigl[\de z /\ell_{\rm N} \bigr]^2+2\eta_{\rm N}\de z /\ell_{\rm N}}
\label{eqn:dx_t_nm} 
\eeq
where $\de z \equiv |\delta x_{\rm b}-\delta x_{\rm ib}|$ is how far the boundary
is from its equilibrium position at the current pressure, with the typical length
\beq
\ell_{\rm N}= \frac{(n_{\rm N}/\chi_{\rm K}^{\rm N})n_{\rm Q}}{2(\gamma-1)g_{\rm b}\ep_{\rm crit}^{\rm N}} \ .
\label{eqn:a_tilde_nm}
\eeq
and all physical quantities are defined similar to those in the $\rm NM \to QM$ half cycle.
\begin{figure*}
\parbox{0.5\hsize}{
\vspace{-2ex}
\includegraphics[width=\hsize]{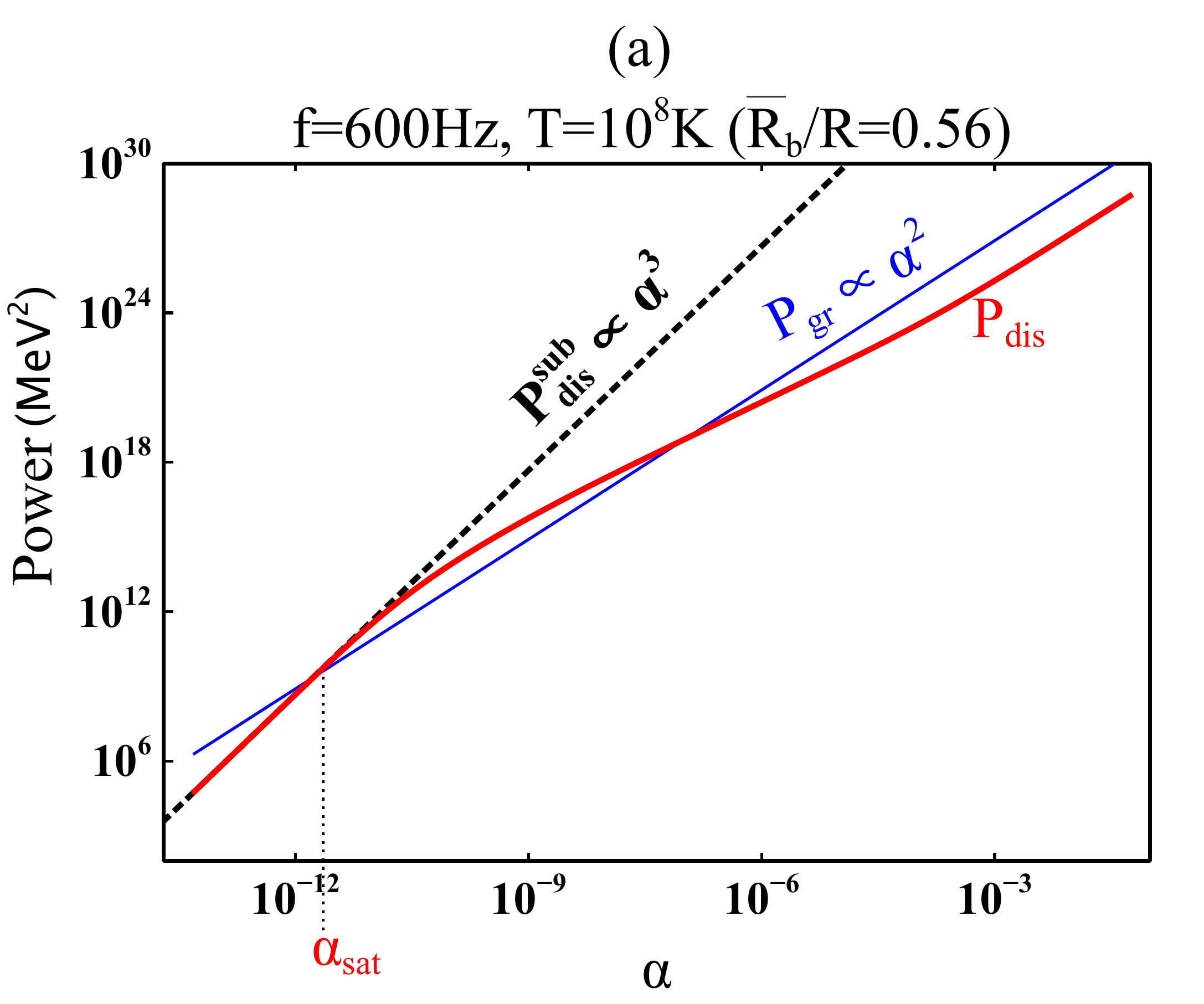}\\[2ex]
}\parbox{0.5\hsize}{
\vspace{-2ex}
\includegraphics[width=\hsize]{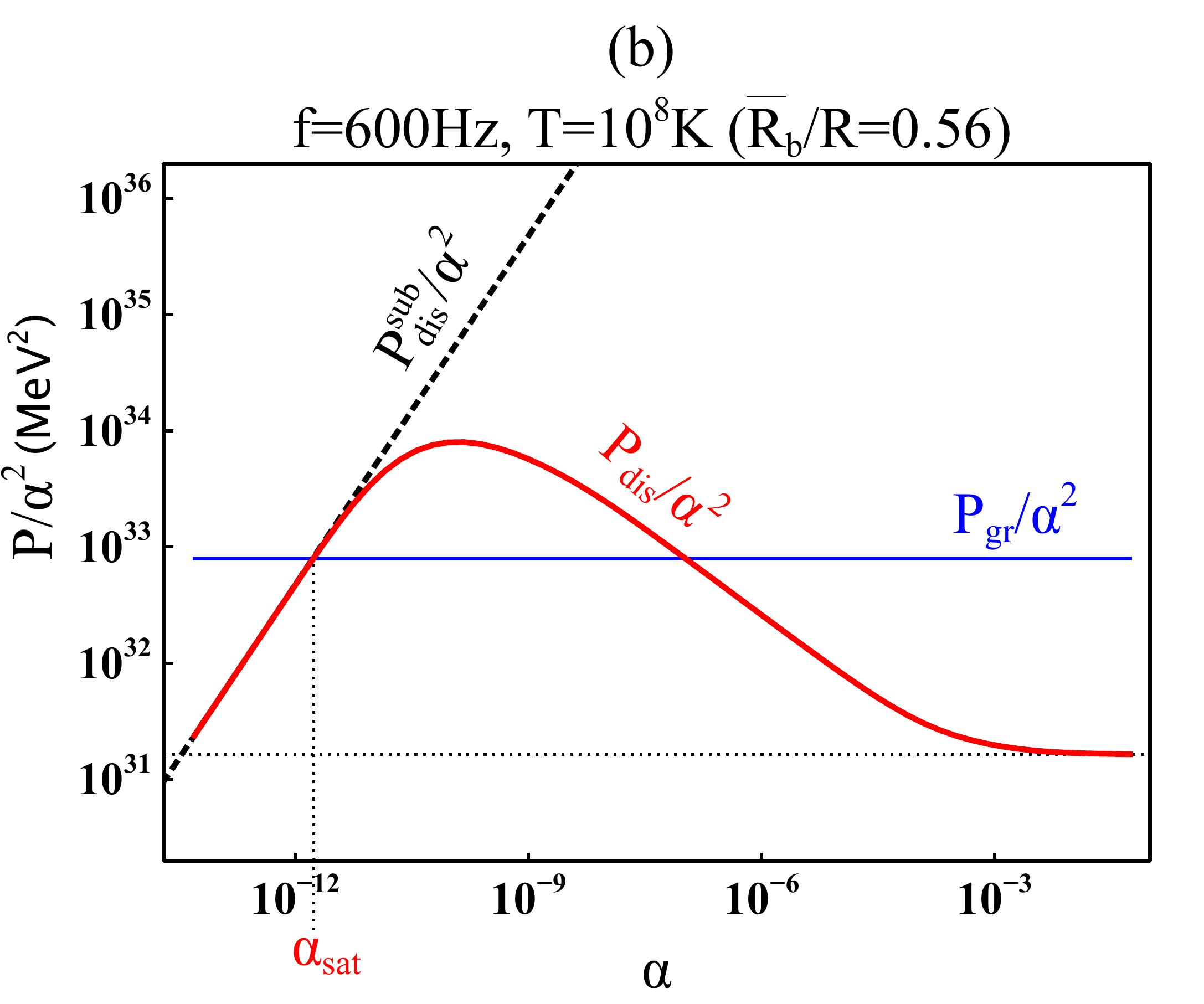}\\[2ex]
}
\caption{(Color online) (a) Dissipated power due to phase conversion $P_{\rm dis}$ (thick solid red curve) as a function of r-mode amplitude $\al$ for a specific example hybrid star (see text). (b) The same quantity where the vertical axis now shows the ratio $P_{\rm dis}/\alpha^2$. At first $P_{\rm dis}$ is proportional to $\alpha^3$ at very low amplitude (dashed line), then at some intermediate amplitude varies less quickly, with a maximum in $P_{\rm dis}/\al^2$, and finally changes to $\al^2$ at higher amplitude.
Also shown is gravitational radiation power $P_{\rm gr}$ (thin solid blue straight line), which is proportional to $\al^2$ at all amplitudes.
The r-mode amplitude will stop growing when dissipation balances
radiation, at the first point of intersection between the two curves.
This defines the saturation amplitude $\alpha_{\rm sat}$.}
\label{fig:P_alpha_complete_HS0}
\end{figure*} 

Therefore with the periodic condition $\delta x_{\rm b}(t)=\delta x_{\rm b}(2\pi/\omega+t)$, Eqs.~\eqn{eqn:dx_t_qm} and \eqn{eqn:dx_t_nm} fully specify the movement of the phase boundary in response to the external pressure oscillation. Next we compute the energy dissipation in this process and see whether it is capable of saturating the r-mode.

\subsection{R-mode damping and saturation amplitude}

During one cycle of an r-mode of amplitude $\al$ the energy dissipated  in a radially oriented cylinder
with an infinitesimal base area $\mathrm{d}S$ straddling the phase
boundary at $(\overline R_{\rm b},\th,\phi)$ is
\beq
 \mathrm{d}W(\alpha, \theta, \phi)=\mathrm{d}S\left(\gamma-1\right)\Delta p_{\rm N} \int_0^\tau\!\!\cos(2\phi+\omega t)\frac{\mathrm{d}\delta R_{\rm b}}{\mathrm{d}t} \mathrm{d}t,
\label{eqn:dW_dp}
\eeq
where the position of the phase boundary $\delta R_{\rm b}(t)$ 
is assumed to move at its maximal speed [see Eqs.~(\ref{eqn:dx_t_qm}) and (\ref{eqn:dx_t_nm})], $\mathrm {d}S=\overline{R}_{\rm b}^2\sin\theta\mathrm {d}\theta\mathrm {d}\phi$ and the pressure oscillation is
\beq
\Delta p_{\rm N}=g_{\rm b}\ep_{\rm crit}^{\rm N}|\delta R_{\rm ib}|=\ep_{\rm crit}^{\rm N}C_{\rm b}\alpha|\sin^2\!\theta\cos \theta| \, 
\label{eqn:dpN_rmode}
\eeq
with $C_{\rm b}=\sqrt{\frac{105}{756\pi}}\Omega ^{2} \overline R_{\rm b}^3/R$  where $\Omega$ and $R$ are the rotation frequency and radius of the star.

Integrating Eq.~\eqn{eqn:dW_dp} over the full range of solid angle gives the total
dissipation of the r-mode in one cycle of oscillation and hence
the total power dissipated, $P_{\rm dis}$.
The r-mode amplitude stops growing (saturates) when this equals
the power injected via back-reaction from gravitational
radiation $P_{\rm gr}$.

As an illustrative example, Fig.~\ref{fig:P_alpha_complete_HS0} 
shows the dissipated power as a function of r-mode amplitude
for a hybrid star rotating with frequency $f=600\,{\rm Hz}$, with 
quark core size $\overline{R}_{\rm b}/R=0.56$
and temperature $T=10^8\,{\rm K}$. For the quark matter EoS
we use the CSS parametrization introduced in Sec.~\ref{sec:CSS-intro}, with $n_{\rm trans}=4\,n_0$, $\Delta \ep/\ep_{\rm trans}=0.2$, and 
$c_{\rm QM}^2=1$. The hadronic matter EoS is taken from 
Ref.~\cite{Hebeler:2010jx}.

In the subthermal regime, the dissipated power first rises with the r-mode amplitude $\al$ as $\al^3$ at very low amplitude, before entering a resonant region with a maximum in $P_{\rm dis}/\al^2$. At high amplitude in the suprathermal regime, the dissipated power is proportional to $\al^2$. The power in gravitational radiation from
the r-mode $P_{\rm gr}$ rises as $\al^2$ at all amplitudes, and is also shown in Fig.~\ref{fig:P_alpha_complete_HS0} for this particular hybrid star.
At low amplitude, the phase conversion dissipation is suppressed
relative to the gravitational radiation and therefore plays no role 
in damping the r-mode. If other damping mechanisms are too
weak to suppress the r-mode, its amplitude will grow. However,
as the amplitude grows, the phase conversion dissipation becomes stronger, and in this example there is a saturation amplitude $\al_{\rm sat}$ at which it equals the gravitational radiation, and the mode stops growing.

Varying parameters such as the size of the quark matter core,
rotation frequency, or temperature of the star will shift the curves in
Fig.~\ref{fig:P_alpha_complete_HS0}, and if the phase conversion dissipation is
too weak then there will be no intersection point ($P_{\rm gr}$ will be
greater than $P_{\rm dis}$ at all $\al$) and phase conversion
dissipation will not stop the growth of the mode.
However, we can see from Fig.~\ref{fig:P_alpha_complete_HS0} that
if saturation occurs, the resultant $\al_{\rm sat}$ is in
the low-amplitude regime, where an analytical approach is available,
and the saturation amplitude is extraordinarily low, of order $10^{-12}$.
This is typical of all model hybrid stars that we investigated.
We derived the analytical expression for the dissipated power 
in the low-amplitude regime (dashed [black] line in Fig.~\ref{fig:P_alpha_complete_HS0}), obtaining
\beq
P_{\rm dis}^{\rm sub}(\alpha)\approx \frac{\alpha^3}{15}\left(\frac{105}{756\pi}\right)^{3/2}\frac{\gamma-1}{\Delta \tilde{p}_{\rm N}}\frac{(\ep_{\rm crit}^{\rm N})^2 \Omega^7 \overline{R}_{\rm b}^{11}}{g_{\rm b} R^3} \ .
\label{eqn:P_sub_approx}
\eeq
where
\beq
\Delta \tilde{p}_{\rm N}\equiv\frac{9}{b_{\rm Q}^2}(\gamma-1)\frac{D_{\rm N}}{\tau_{\rm N}}\frac{\eta_{\rm N}\left(g_{\rm b}\ep_{\rm crit}^{\rm N}\right)^2}{n_{\rm Q}\left(n_{\rm N}/\chi_{\rm K}^{\rm N}\right)}\frac{1}{\Omega^2}. 
\label{eqn:dp_nm_tilde}
\eeq
For the class of models we have analyzed, in general because in nuclear matter diffusion is less efficient ($D_{\rm N}/D_{\rm Q}\approx O(10^{-2})$) and weak interactions take more time to proceed 
($\tau_{\rm N}/\tau_{\rm Q}\approx O(10^2)$), therefore the $\rm {QM \to NM}$ transition half cycle dominates the dissipation.

Eq.~\eqn{eqn:P_sub_approx} allows us to assess how the strength of
phase conversion dissipation depends
on the various parameters involved.
It is particularly sensitive to the size of the quark core, 
and this will be important when
considering a whole family of hybrid stars with different
central pressures and hence different core sizes.

\begin{figure}[htb]
\includegraphics[width=\hsize]{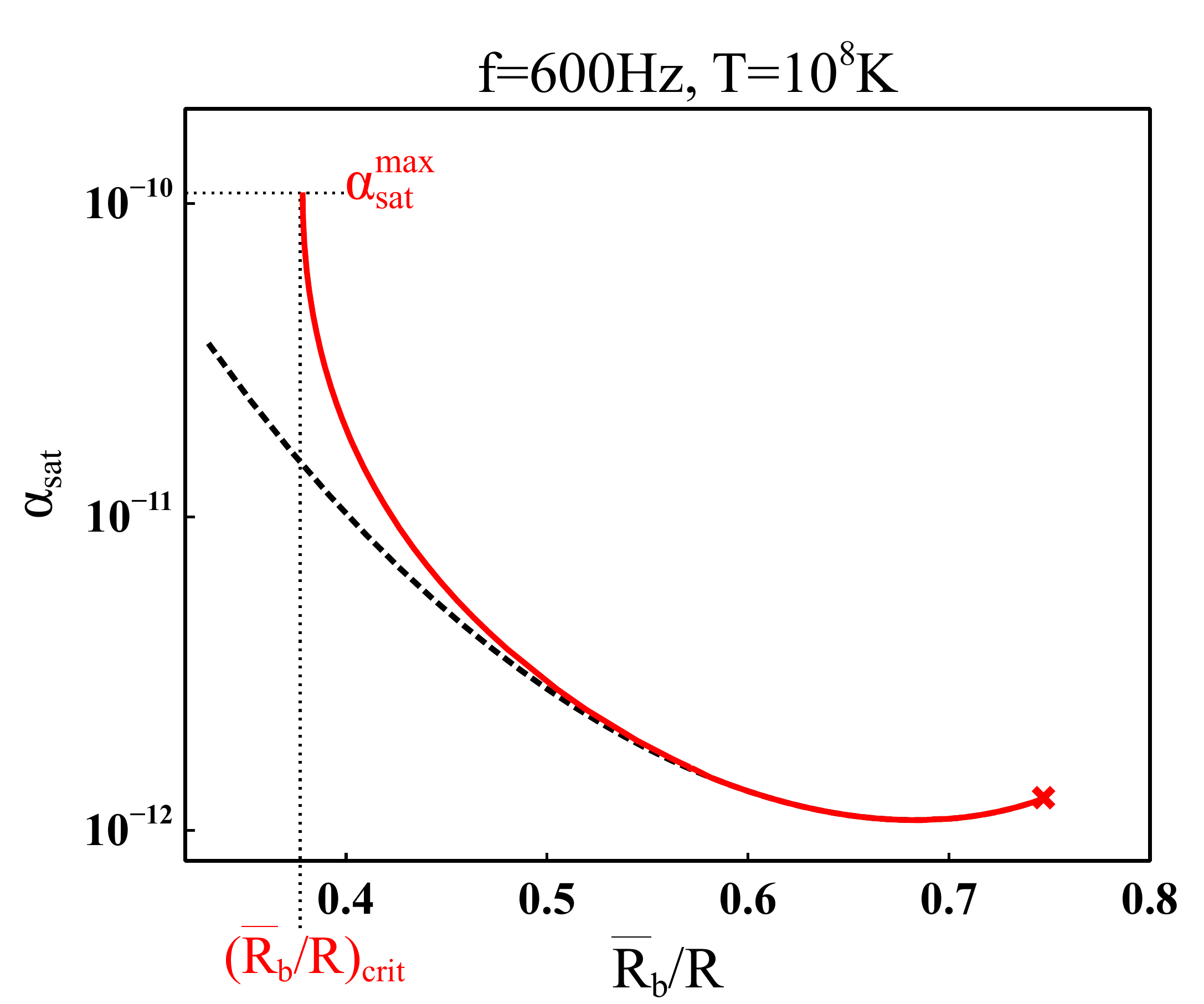}
\caption{(Color online) R-mode saturation amplitude (red solid curve) and its low-amplitude analytical approximation (black dashed curve) as a function of the 
radius of the quark matter core $\overline{R}_{\rm b}$ divided by the star radius $R$ in a family of hybrid stars.
For $\overline{R}_{\rm b}/R < (\overline{R}_{\rm b}/R)_{\rm crit}\approx0.38$, damping is too weak
to saturate the r-mode. At $\overline{R}_{\rm b}/R \gtrsim 0.75$ the
hybrid star is unstable against gravitational collapse. The mass fraction of the core is in the range $0.12\lesssim M_{\rm core}/M_{\rm star}\lesssim0.68$ for all configurations shown on the red solid curve.}
\label{fig:alpha_sat_core_ratio_HS0_600Hz}
\end{figure}

The results of such an investigation are shown in
Fig.~\ref{fig:alpha_sat_core_ratio_HS0_600Hz}, where the solid (red) curve
gives the numerically calculated saturation amplitude
($\alpha_{\rm sat}$ in Fig.~\ref{fig:P_alpha_complete_HS0}) as a function of
the size of the quark matter core in units of the star radius,
$\overline{R}_{\rm b}/R$. To construct this curve we used
the hadronic and quark matter EoSs of
Fig.~\ref{fig:P_alpha_complete_HS0} and varied the central
pressure, yielding a family of different star configurations. As
$\overline{R}_{\rm b}/R$ decreases, the dissipation power $P_{\rm dis}$ 
decreases rapidly relative to the gravitational radiation $P_{\rm gr}$.
The relative shift in the two corresponding curves in
Fig.~\ref{fig:P_alpha_complete_HS0} leads to
an upper limit on $\alpha_{\rm sat}$ when
$P_{\rm gr}$ is tangent to $P_{\rm dis}$.
This corresponds to the end of the solid curve 
in Fig.~\ref{fig:alpha_sat_core_ratio_HS0_600Hz} at
$\alpha_{\rm sat}^{\rm max}$ at the
critical value of the quark core size, $(\overline{R}_{\rm b}/R)_{\rm crit}$, below
which the phase conversion mechanism cannot saturate the r-mode any more.

The black dashed curve in
Fig.~\ref{fig:alpha_sat_core_ratio_HS0_600Hz} is the low-amplitude analytical approximation to $\al_{\rm sat}$ 
\bea
&&\alpha_{\rm sat}^{\rm approx}=\left(\frac{2^{22}\pi^{9/2}}{3^3\cdot5^{5/2}} \right)\!G\, \frac{\tilde{D}_{\rm N}}{\tau _{\rm N}} \frac{\left(\chi_{\rm K}^{\rm N}\right)^3}{n_{\rm Q}n_{\rm N}^3 b_{\rm Q}^2} \frac{g_{\rm b}^3 M^2 \tilde{J}^2}{\Omega} \frac{R^9}{\overline{R}_{\rm b}^{11}}\nonumber \\[2ex]
&&\approx 4.2\times 10^{-11}\;\gamma \left(\frac{\tilde{D}_{\rm N}}{1.5\,{\rm MeV}}\right)\left(\frac{\tau_{\rm N}}{2\times 10^{-8}\,{\rm s}}\right)^{-1}\nonumber \\
&&\times\  \left(\frac{b_{\rm Q}}{1/3}\right)^{-2} \left(\frac{n_{\rm N}}{2\,n_0}\right)^{-4} \left(\frac{\chi_{\rm K}^{\rm N}}{(100\,{\rm MeV})^2}\right)^{3}\left(\frac{g_{\rm b}}{g_{\rm u}}\right)^{3}\nonumber \\
&&\times \left(\frac{\ep_{\rm crit}^{\rm Q}}{2\,\ep_{\rm crit}^{\rm N}}\right)^{3}\left(\frac{\ep_{\rm crit}^{\rm N}}{600\,{\rm MeV\,fm^{-3}}}\right)^{3}  \left(\frac{M}{1.4\,M_{\odot}}\right)^{2}\nonumber \\
&&\times  \left(\frac{\tilde{J}}{0.02}\right)^2 \left(\frac{f}{1 {\rm kHz}}\right)^{-1} \left(\frac{R}{10\,{\rm km}}\right) \left(\frac{\overline{R}_{\rm b}/R}{0.4}\right)^{-8},
\label{eqn:alpha_sat_analytic}
\eea
where $D_{\rm N}\equiv \tilde{D}_{\rm N}\times T^{-2}$ 
, and $g_{\rm u}$ is the Newtonian gravitational acceleration at the phase boundary when the quark core has uniform density $\ep=\ep_{\rm crit}^{\rm Q}$ ($g_{\rm u}\equiv \frac{4}{3}\pi G \ep_{\rm crit}^{\rm Q} \overline{R}_{\rm b}$).
This approximation is very accurate when the phase conversion damping
is strong, but it does not capture the sudden weakening of that dissipation
when, for example, the core radius becomes small. 

After exploring more CSS parameter space [$n_{\rm trans}$, $\Delta \ep/\ep_{\rm trans}$, $c_{\rm QM}^2$], we find that in the phase conversion mechanism, the most sensitive quark matter EoS parameter is the transition pressure (for example when the mass of the hybrid star is fixed), and the higher it is the larger r-mode saturation amplitude is predicted.

\subsection{Range of validity of low-amplitude approximation}

The range of validity of 
Eqs.~(\ref{eqn:P_sub_approx}) and (\ref{eqn:alpha_sat_analytic})
is found by calculating the next-to-leading (NLO) contribution, and requiring that it be less than a fraction $\epsilon$ of the total dissipated power.
We find 
that the approximation is valid when
\bea
&& \epsilon \geqslant  \frac{2^{39}\pi^5}{3^{12}5^4} \frac{G^2}{(\gamma-1)^2}\frac{g_{\rm b}^2 M^4 \tilde{J}^4\Omega^6}{\left(\ep_{\rm crit}^{\rm N}\right)^2}\left(\frac{R}{\overline{R}_{\rm b}}\right)^{16}\nonumber \\
&&\simeq 2.96 \left(\frac{\gamma-1}{0.5}\right)^{-2} \left(\frac{\ep_{\rm crit}^{\rm Q}}{2\,\ep_{\rm crit}^{\rm N}}\right)^{2} \left(\frac{g_{\rm b}}{g_{\rm u}}\right)^{2} \left(\frac{M}{1.4\,M_{\odot}}\right)^{4} \nonumber \\
&&\times \left(\frac{\tilde{J}}{0.02}\right)^{4} \left(\frac{f}{1\,\rm{kHz}}\right)^{6}\left(\frac{R}{10\,{\rm km}}\right)^{2}\left(\frac{\overline{R}_{\rm b}/R}{0.4}\right)^{-14} \!\!  .
\label{eqn:ep_astro}
\eea
We see that the validity of the low-amplitude approximation is
mainly determined by the size of the quark matter core and the
rotation frequency of the star.

\subsection{Discussion}
We have described how phase conversion
in a multicomponent compact star provides a mechanism for
damping density oscillations, via the 
phase lag in the response of the interface between components
of different baryon densities to the applied pressure oscillation.
The phase lag arises from the finite rate of interconversion between
the phases, which limits the speed with which the interface can move.
We studied the case where the two phases are separated by a 
sharp boundary (first-order phase transition)
and analyzed the movement of the interface in the approximation
of a steady state, neglecting additional acceleration effects and
complicated hydrodynamic effects like
turbulence. In particular, we studied the astrophysically
interesting case of the damping of r-mode oscillations
\cite{Andersson:1997xt,Andersson:2000mf} in a two-component star.
We found that  phase conversion dissipation does not affect the
r-mode instability region, because it vanishes as $\al^3$ at low
r-mode amplitude $\al$. 
However, depending on the values of relevant parameters,
phase conversion dissipation can either saturate the r-mode at extremely low amplitudes,
$\al_{\rm sat} \lesssim 10^{-10}$ 
in the explicit example of hadron-quark transformation
at the sharp quark-hadron interface in a hybrid star, or be insufficient to saturate the r-mode at all.
The reason for this behavior stems, analogously to the bulk viscosity \cite{Alford:2010gw}, from the resonant character of the dissipation, which is relatively strong when the time scale of the dissipation matches the time scale of the external oscillation (see Fig.~\ref{fig:P_alpha_complete_HS0}). Whether saturation is possible depends therefore on the microscopic and astrophysical parameters, like in particular on the mass of the quark core, which should not be too small.

Our main result is Eq.~\eqn{eqn:dW_dp}, which must be evaluated using numerical solutions of Eqs.~\eqn{eqn:dx_t_qm} and \eqn{eqn:dx_t_nm}. We also give the low-amplitude analytic expressions for the power dissipated
[Eq.~\eqn{eqn:P_sub_approx}] and the saturation amplitude 
[Eq.~\eqn{eqn:alpha_sat_analytic}] which are valid
when the dissipation is sufficiently strong, obeying Eq.~\eqn{eqn:ep_astro}
with $\epsilon \ll 1$.

Our results have significant implications for astrophysical
signatures of exotic high-density phases of matter, such as quark matter.
The observed data for  millisecond pulsars is not consistent with
the minimal model of pulsars as stars made of nuclear matter
with damping of r-modes
via bulk and shear viscosity \cite{Alford:2013pma}. Resolving this discrepancy
requires either a new mechanism for stabilizing r-modes, or
a new mechanism for saturating unstable r-modes at 
$\alpha_{\rm sat} \lesssim 10^{-8} \!\!-\!\!10^{-7}$ \cite{Alford:2013pma,Mahmoodifar:2013quw,Haskell:2012vg}. Previously proposed mechanisms have problems to achieve this.
Suprathermal bulk viscosity and hydrodynamic oscillations 
both give $\al_{\rm sat}\sim 1$ \cite{Alford:2011pi,Lindblom:2000az}.
The nonlinear coupling of the r-mode to viscously
damped daughter modes could give $\alpha_{\rm sat} \sim 10^{-6}$ to $10^{-3}$ 
\cite{Brink:2004bg,Bondarescu:2013xwa}. 
The recently proposed vortex-fluxtube cutting mechanism
\cite{Haskell:2013hja} might give sufficiently small 
saturation amplitudes but is present only at sufficiently low temperatures $T \ll T_{c}\lesssim 10^9$ K, which could be exceeded by the r-mode (and/or accretion) heating \cite{Alford:2013pma}.
One of the main results of this paper is that
phase conversion dissipation can provide saturation at the required
amplitude to explain millisecond pulsar data.

Second, due to the extremely low r-mode saturation amplitude of our proposed mechanism, hybrid stars would behave very differently from neutron or strange stars. As discussed in Ref.~\cite{Alford:2013pma}, if the known millisecond sources were hybrid stars then, for the low saturation amplitudes that we have found, they would have cooled out of the r-mode instability region quickly (in millions of years) so that they would have very low temperatures by now. In contrast, in neutron stars r-modes would be present and would provide such strong heating that the temperature of observed millisecond pulsars would be $T_\infty \sim O(10^5\!\!-\!\!10^6)$ K \cite{Alford:2013pma}. This prediction assumes a (so far unknown) saturation mechanism that would saturate the mode at a value $\alpha_{\rm sat} \lesssim 10^{-8}$ required by the pulsar data. This temperature is significantly higher than what standard cooling estimates suggest for such old sources. The same holds for strange quark stars where the enhanced viscous damping can explain the pulsar data, but even in this case the star would spin down along the boundary of the corresponding stability window which would keep it at similarly high temperatures. Measurements of or bounds on temperatures of isolated millisecond pulsars provide therefore a promising way to discriminate hybrid stars.

\section{Summary and outlook}
\label{sec:summary}

In this paper we studied how strange quark matter could influence the static properties and dynamic behaviors of hybrid stars in two ways.

Firstly we explored the CSS (Constant-Speed-of-Sound) parameterization
of the quark matter EoS, which assumes
a sharp transition from nuclear maftter to quark
matter, and that the speed of sound in quark matter is independent of the
pressure.
Our study is intended to motivate the use of the CSS parameterization as a framework in which the implications of observations of neutron stars for the high-density EoS can be expressed and discussed in a way that is reasonably model-independent.
We showed how mass and radius observations can be expressed
as constraints on the three CSS parameters, and we found that the observation of a $2.0\,\Msolar$ star already
constrains the CSS parameters significantly.

If, as predicted by many models, $\cQMsq\lesssim 1/3$,
then there are two possible scenarios: a low-transition-pressure scenario, where
the transition to the high density phase occurs at
$\ntrans \lesssim 2\,n_0$, and a high-transition-pressure scenario.
In the low-transiti-
on-pressure case there are strong
constraints on the radius of the star, as shown in
Fig.~\ref{fig:CSS-radius-zoom}.  The radius of the maximum mass
star (which is typically the smallest possible star) must be great-
er
than about 11.25\,km, and the radius of a $1.4\,\Msolar$
star must be greater than about 12\,km \cite{Lattimer:2012nd}.
The high-transition-pressure case corresponds
to the white region on the
right side of the left panels of Figs.~\ref{fig:CSS-max-mass},~\ref{fig:CSS-radius-max-mass},~\ref{fig:CSS-radius}, which tends to give a very small branch of hybrid stars
with tiny quark matter cores,
occurring in a narrow range of central pressures just above the
transition pressure. 

If $\cQMsq$ is larger than $1/3$ then a larger region of the CSS parameter
space becomes allowed. The right panels of
Figs.~\ref{fig:CSS-max-mass},~\ref{fig:CSS-radius-max-mass},~\ref{fig:CSS-radius}
show the extreme case where $\cQMsq=1$. In this case the minimum possible
radius is around 9.0\,km, and the radius of a $1.4\,\Msolar$ star must be
greater than about 9.5\,km. 

In the second part of the paper we studied a dissipation mechanism
that is expected to occur in hybrid stars with a sharp boundary between 
the quark matter and the nuclear matter. The mechanism is phase conversion 
dissipation, and we found that it can saturate an unstable r-mode 
with an amplitude in the range $O(10^{-12})$ to $O(10^{-10})$.
This is of the right order to explain millisecond pulsar data. We gave an analytical form of the saturation amplitude, with its dependence on a set of parameters such as quark core size, rotation frequency and so on.
More measurements of temperatures would give more information about the r-mode amplitude and therefore help probe the compositions of hybrid stars.

As well as the quark-hadron interface in a hybrid star, any first-order phase transition that leads to a sharp interface between two phases with different baryon densities could lead to phase conversion dissipation of global pressure oscillation modes. For example, such dissipation could occur
at the boundary between different phases of quark matter; for the transition from nuclei to nucleon, as long as there is a difference of baryon densities the mechanism will operate as well.

Our discussion was limited to the case of a sharp interface, which is the expected configuration if the surface tension is large enough. If the surface tension is small, there will instead be a mixed-phase region where domains of charged hadronic and quark matter coexist. We expect that the phase conversion dissipation mechanism will operate in this case too, as the domains expand and shrink in response to pressure oscillations. A similar mechanism should also be relevant for the ``nuclear pasta" mixed phases in the inner crust of an ordinary neutron star. In this case in addition to the slow beta-equilibration processes there may also be slow strong interaction equilibration processes, whose rate is suppressed by tunneling factors for the transition between geometric domains of different size. This could further enhance the dissipation.

\section{Acknowledgments}
This work was supported by the U.S.\ Department of Energy,
Office of Science, Office of Nuclear Physics under Award Number
\#DE-FG02-05ER41375, and by the DOE Topical Collaboration ``Neutrinos
and Nucleosynthesis in Hot and Dense Matter'' contract \#DE-SC0004955.

\newcommand{\apjl}{Astrophys. J. Lett.\ }
\newcommand{\mnras}{Mon. Not. R. Astron. Soc.\ }
\newcommand{\aap}{Astron. Astrophys.\ }

\bibliographystyle{JHEP_MGA}
\bibliography{EPJA} 

\end{document}